\title{Partial-MDS Codes and their Application to RAID Type of
Architectures}
\author{Mario Blaum, James Lee Hafner and Steven Hetzler\\
IBM Almaden Research Center\\
San Jose, CA 95120
}
\documentclass[12pt]{article}
\usepackage{latexsym}
\usepackage{multirow}
 \newtheorem{theo}{Theorem}[section]
 \newtheorem{lemma}{Lemma}[section]
 \newtheorem{constr}{Construction}[section]
 \newtheorem{defin}{Definition}[section]
 \newtheorem{ex}{Example}[section]

\newtheorem{COROLLARY}{\indent Corollary}

\newtheorem{EXAMPLE}{\indent Example}

\newtheorem{THEOREM}{\indent Theorem}

\newtheorem{REMARK}{\indent Remark}

\newcommand{\ga}{\mbox{$\gamma$}}

\newcommand{\fullstop}{\hspace{-0.85em} {\bf .}}

\newcommand{\uc}{\mbox{$\underline{c}$}}

\newcommand{\ra}{\mbox{$\rightarrow$}}

\newcommand{\la}{\mbox{$\leftarrow$}}

\newcommand{\al}{\mbox{$\alpha$}}

\newcommand{\eq}{\mbox{$\, =\,$}}

\newcommand{\qed}{\hfill$\Box$\\[1ex]}
\newcommand{\pf}{{\bf Proof: }}

\newcommand{\uw}{\mbox{$\underline{w}$}}
\newcommand{\uzero}{\mbox{$\underline{0}$}}

\newcommand{\xor}{\mbox{$\,\oplus\,$}}

\newcommand{\C}{\mbox{${\cal C}$}}

\newcommand{\cH}{\mbox{${\cal H}$}}

\newcommand{\br}{\\ }
\newcommand{\ce}{\begin{center}}
\newcommand{\cen}{\end{center}}

\newcommand{\ipb}{\begin{description}}
\newcommand{\ipn}{\end{description}}

\newcommand{\qb}{\begin{quote}}
\newcommand{\qn}{\end{quote}}

\newcommand{\tp}{\begin{titlepage}}
\newcommand{\tpn}{\end{titlepage}}
\newcommand{\zb}{\begin{figure}[hbtp]}
\newcommand{\zn}{\end{figure}}

\newcommand{\EQX}[1]{\begin{equation}\label{#1}}
\newcommand{\ENX}{\end{equation}}
\newcommand{\EQL}{\begin{eqnarray*}}
\newcommand{\ENL}{\end{eqnarray*}}
\newcommand{\EQLX}[1]{\begin{eqnarray}\label{#1}}
\newcommand{\ENLX}{\end{eqnarray}}

\newcommand{\open}{\begin{document}}
\newcommand{\close}{\end{document}}
\newcommand{\lfcr}[1]{\br\hspace*{#1em}}
\newenvironment{mat}[1]
{\left[\begin{array}{#1}}{\end{array}\right]}
\newcommand{\GAMMA}{\Gamma}
\newcommand{\DELTA}{\Delta}
\newcommand{\THETA}{\Theta}
\newcommand{\LAMBDA}{\Lambda}
\newcommand{\XI}{\Xi}
\newcommand{\PI}{\Pi}
\newcommand{\SIGMA}{\Sigma}
\newcommand{\UPSILON}{\Upsilon}
\newcommand{\PHI}{\Phi}
\newcommand{\PSI}{\Psi}
\newcommand{\OMEGA}{\Omega}
\newcommand{\bldgreek}[1]{\mbox{\boldmath $#1$}}
\newcommand{\bldbeta}{\bldgreek{\beta}}
\newcommand{\bldgamma}{\bldgreek{\gamma}}
\newcommand{\blddelta}{\bldgreek{\delta}}
\newcommand{\bldepsilon}{\bldgreek{\epsilon}}
\newcommand{\bldvarepsilon}{\bldgreek{\varepsilon}}
\newcommand{\bldzeta}{\bldgreek{\zeta}}
\newcommand{\bldeta}{\bldgreek{\eta}}
\newcommand{\bldtheta}{\bldgreek{\theta}}
\newcommand{\bldvartheta}{\bldgreek{\vartheta}}
\newcommand{\bldiota}{\bldgreek{\iota}}
\newcommand{\bldkappa}{\bldgreek{\kappa}}
\newcommand{\bldlambda}{\bldgreek{\lambda}}
\newcommand{\bldmu}{\bldgreek{\mu}}
\newcommand{\bldnu}{\bldgreek{\nu}}
\newcommand{\bldxi}{\bldgreek{\xi}}
\newcommand{\bldpi}{\bldgreek{\pi}}
\newcommand{\bldvarpi}{\bldgreek{\varpi}}
\newcommand{\bldrho}{\bldgreek{\rho}}
\newcommand{\bldvarrho}{\bldgreek{\varrho}}
\newcommand{\bldsigma}{\bldgreek{\sigma}}
\newcommand{\bldvarsigma}{\bldgreek{\varsigma}}
\newcommand{\bldtau}{\bldgreek{\tau}}
\newcommand{\bldupsilon}{\bldgreek{\upsilon}}
\newcommand{\bldphi}{\bldgreek{\phi}}
\newcommand{\bldvarphi}{\bldgreek{\varphi}}
\newcommand{\bldchi}{\bldgreek{\chi}}
\newcommand{\bldpsi}{\bldgreek{\psi}}
\newcommand{\bldomega}{\bldgreek{\omega}}
\begin{document}
\parindent=10pt
\maketitle
%%%%%%%%%%%%%%%%%%%%%%%%%%%%%%%%%%%%%%%%%%%%%%%%%%%%%%%%%%%%%%%%%%%%%%%%%%%
%%
\begin{abstract}
A family of codes with a natural two-dimensional
structure is presented, inspired by an application of RAID type of
architectures whose units are
solid state drives (SSDs). Arrays of SSDs behave
differently to arrays of hard disk drives (HDDs), since hard errors
in sectors are common and traditional RAID approaches (like RAID 5 or
RAID 6) may be either insufficient or excessive. An efficient
solution to this problem is given by the new codes presented, called
partial-MDS (PMDS) codes.

\vspace{.3cm}

\end{abstract}

\section{Introduction}
\label{Introduction}
Consider an array of, say, $n$ storage devices. Each storage device
contains a (large) number of sectors, each sector protected by an
error-correcting code (ECC) dealing with the most common
errors in the media. However, it may occur that one or more
of the storage devices experiences a catastrophic failure. In that
case, data loss will occur if no
further protection is implemented. For that reason, the architecture
known as Redundant Arrays of Inexpensive Disks (RAID) was
proposed~\cite{g}.

The way RAID architectures work is by assigning one or more devices
to parity. For instance, take $n$ sectors in the same location in
each device (we call this set of $n$ sectors, a ``stripe''): $n-1$
sectors carry information, while the $n$th is the XOR of the $n-1$
information sectors. We repeat this for each stripe of sectors in the
array. Such an architecture is called a RAID 4 or a RAID 5 type of
architecture. In what follows, we will call it RAID 5, the difference
between RAID 5 and RAID 4 consisting on the distribution of the
parity sectors, but we do not address this issue here.
A RAID 6 architecture gives protection against two catastrophic
failures. 

From a coding point of view, the model of failures corresponds to
erasures, i.e., errors whose location is known~\cite{lc}\cite{lc2}.
It is preferrable to use Maximum Distance Separable (MDS) codes for
RAID 6 types of architectures: in order to correct two erasures 
exactly two parities are needed. There are many choices for MDS codes
correcting two or more erasures: we can use Reed-Solomon (RS)
codes~\cite{ms}\cite{p}, or array codes, like EVENODD~\cite{bbbm}, 
RDP~\cite{Cor+04}, X-codes~\cite{xb}, B-codes~\cite{xbbw},
C-codes~\cite{ls}, %HoVer codes~\cite{h}, 
Liberation codes~\cite{p2}, %WEAVER codes~\cite{h2}, 
and others~\cite{bft}\cite{tb}. 
%Under certain conditions that we do
%not address here, it may be 

Architectures like RAID 5 and RAID 6 are efficient when the storage
devices are hard disk drives (HDDs). However, when using solid state
drives (SSDs) like flash, these types of architectures, by themselves,
either are not efficient or they are wasteful. Arrays of SSDs pose
new challenges for code design, so we will spend the rest of this
section addressing some of them. Different ways to adapt RAID
architectures to SSDs are being considered in recent literature. For
instance, ways to enhance the performance of RAID 5 are described
in~\cite{bkpm}\cite{is}. See also~\cite{glmsw}\cite{hu}, where the
internal ECC and a RAID type of architecture communicate. In
particular, \cite{glmsw} uses an adaptive method to increase the
redundancy when the bit-error rate increases.

Contrary to HDDs, SSDs degrade significantly in time and as a
function of the number 
of writes~\cite{M}. As time goes by and the number of writes
increases, the likelihood of a hard error in a sector also increases.
A hard error occurs mainly when the internal ECC of a sector is exceeded. In
general, BCH codes~\cite{lc}\cite{lc2} are used for the internal ECC
of a sector, although many other
codes (including LDPC codes with soft decoding) are possible.
Moreover, in recent years some remarkable non-traditional approaches for the ECC
that exploit the assymetry of the SSD channel have been
developed~\cite{bam}\cite{hla}\cite{jbb}\cite{jmsb}\cite{jsb}\cite{kbe}\cite{klny}\cite{msvwy}.
However, we do not address 
the internal ECC problem in this paper. The point is, a hard error
corresponds to an uncorrectable error in a sector. Normally, the ECC
is coupled with a Cyclic Redundancy Code (CRC), which detects the
situation when the ECC miscorrects (the ECC has an inherent detection
capability that may not be sufficient, hence it often needs to
be reinforced by the CRC). There are several ways to implement the CRC,
but we do not address them here. We will assume instead that a
hard error means that the information in a sector is lost (an erased
sector) and that we can detect this situation.

From the discussion above, we see that, contrary to arrays of HDDs,
arrays of SSDs present a mixed failure mode: on one hand we have
catastrophic SSDs failures, as in the case of HDDs. On the other
hand, we also have hard errors, which in general are silent: 
their existence is unknown until the sectors are accessed. This
situation complicates the task of a RAID type of architecture. In
effect, assume that a catastrophic SSD failure occurs in a RAID 5
architecture. Each sector of the failed
device is reconstructed by XORing the corresponding sectors in each
stripe of the surviving 
devices. However, if there is a stripe that in addition has
suffered a 
hard error, such a stripe has two sectors that have failed. Since we are
using RAID 5, we cannot recover from such an event and data loss
will occur.

A possible solution to the situation above is using a RAID 6 type
of architecture, in which two SSDs are used for parity. Certainly, this
architecture allows for the recovery of two erased sectors in the
same stripe. 
However, such a solution is expensive, since it requires an additional
whole device to protect against hard errors. Moreover, two hard
errors in a stripe, in addition to the catastrophic device
failure, would still cause data loss, and such a
scenario may not be unlikely, depending on the statistics of errors.
We would like some 
solution intermediate between RAID 5 and RAID 6 allowing the
handling of hard errors without the need of dedicating a whole second
SSD to parity, and in addition being able to handle at least two hard
errors in the same stripe, a catastrophic failure having occurred. 

In order to handle this mixed environment of hard errors with
catastrophic failures, we need to take into account the way
information is written in SSDs, which is quite different to the way
it is done in HDDs. In an SSD, a new write consists of erasing
first a number of consecutive sectors and then rewriting all of them.
Therefore, the short write operation in arrays of SSDs (like one
sector at a time) is not an
issue here: each time a new write is done, a group of, say, $m$
sectors in each SSD is erased and then rewritten. So, the parity
needs to be recomputed as part of the new write. We can assume that
the array consists of
$m\times n$ blocks (i.e., each block consists of $m$ stripes),
repeated one after the other. Each $m\times n$ 
block is an independent unit, and we will show how to compute the
parity for each block. Also, each new write consists of writing
a number of $m\times n$ blocks (this number may be one, depending on
the application, the particular SSD used, and other factors). Our
goal is to present a family of codes, that we call partial-MDS (PMDS)
codes, allowing for the simultaneous correction of catastrophic failures and
hard errors. 

The paper is organized as follows: in Section~\ref{PMDS}, we present
the theoretical framework as well as the 
basic definitions. In 
Section~\ref{construction}, we present our main construction. In
Section~\ref{special}, we study the special case in which the general
construction of Section~\ref{construction} extends RAID 5, and we
find the general conditions for such codes to be PMDS. 
In Section~\ref{scases}, we study specific cases with parameters
relevant to applications, each case analyzed in a separate
subsection. In Section~\ref{salt}, we present an alternative
construction to the one presented in Section~\ref{construction}, we
compare the two and we study some relevant special cases of this
second construction. In Section~\ref{simple} we present a third
construction for cases extending RAID 5. This third construction is
not as powerful as the the previous ones (it cannot handle three
erasures in the same stripe) but uses finite fields of smaller size,
simplifying the implementation. 
In Section~\ref{prob}, we compute the probability of data loss when a
catastrophic device failure has occurred under different scenarios.
We conclude the paper by drawing some conclusions. 

Although the results can be extended to finite fields of arbitrary
characteristic, for simplicity, we consider only fields of
characteristic 2. 

\section{Partial-MDS codes}
\label{PMDS}

Consider an $m\times n$ array, each entry of the array consisting of
$b$ symbols (we assume that each of the $b$ symbols is a bit for the
sake of the description, but in practice it may be a much larger
symbol). Each stripe in the array is protected by
$r$ parity entries in such a way that any $r$ erasures in the stripe
will be recovered. In other words, each stripe of the array constitutes
and $[n,n-r,r+1]$ MDS code. In addition, we will add $s$ extra ``global''
parities. Those $s$ extra parities may be placed in different
ways in the array, but in order to simplify the description we will
place them in the last stripe. Being global means that these parities
affect all $mn$ entries in the array. For instance, Figure~\ref{fig1}
shows a $4\times 5$ array with $r\eq 1$ and $s\eq 2$ such that the
two extra global parities are placed in the last stripe.

\begin{figure}
$$\begin{array}{|c|c|c|c|c|}
\hline
\phantom{P}&&&&P\\
\hline
&\phantom{P}&&&P\\
\hline
&&&&P\\
\hline
&&P&P&P\\
\hline
\end{array}$$
\caption{A $4\times 5$ array with $r\eq 1$ and $s\eq 2$}
\label{fig1}
\end{figure}

The idea of a partial-MDS code (to be defined formally), is the
following: looking at Figure~\ref{fig1}, assume that a catastrophic
failure occurs (that is, a whole column in the array has failed), and
in addition, we have up to two hard errors anywhere in the array.
Then we want the code to correct
these failures (erasures in coding parlance). The situation is
illustrated in Figure~\ref{fig2}, where the hard errors are indicated
with the letter `H': the two hard errors may occur
either in different stripes or in the same stripe.

\begin{figure}
$$\begin{array}{cc}
\begin{array}
{|c|c|c|c|c|}
\hline
\phantom{P}&H&\phantom{P}&F&\\
\hline
&\phantom{P}&&F&\\
\hline
&&&F&H\\
\hline
&&&F&\\
\hline
\end{array}
&
\begin{array}
{|c|c|c|c|c|}
\hline
\phantom{P}&&\phantom{P}&F&\\
\hline
&\phantom{P}&&F&\\
\hline
H&&&F&H\\
\hline
&&&F&\\
\hline
\end{array}
\end{array}
$$
\caption{$4\times 5$ arrays with a catastrophic failure and two hard
errors}
\label{fig2}
\end{figure}

A natural way of solving this problem is by using an MDS code. In our
$4\times 5$ array example, we have a total of 6 parity sectors. So,
it is feasible to implement an MDS code on $20$ symbols with 6 parity
symbols. In other words, a $[20,14,7]$ MDS code (like a RS code). The
problem with
this approach is its complexity. The case of a $4\times 5$ array is given
for the purpose of illustration, but more typical values of $m$ in
applications are $m\eq 16$ and even $m\eq 32$. That would give 18 or
34 parity sectors. Implementing such a code, although feasible, is
complex. We want that the code, in normal operation, utilizes its
underlying RAID structure based on stripes, like single parity in the
case of RAID 5. The extra parities are invoked in rare occasions.
So, given this constraint of an horizontal code, we want to establish
an optimality criterium for codes, that we will call partial-MDS (PMDS)
codes. In the case of the example of RAID 5 plus two global parities,
we want the code to correct up to one erasure per stripe, and in
addition, two extra erasures anywhere. For example, the code of
Figure~\ref{fig1} is PMDS if it can correct any of the situations
depicted in Figure~\ref{fig2}. Formally,

\begin{defin}
\label{defPMDS}
{\em
Let $\C$ be a linear $[mn,m(n-r)-s]$ code over a field or ring such
that when codewords are taken row-wise as 
$m\times n$ arrays, each row belongs in an
$[n,n-r,r+1]$ MDS code.
Given $(s_1,s_2,\ldots,s_t)$ such that each $s_j\geq 1$ and
$\sum_{j=1}^ts_j\eq s$, we say that 
$\C$ is $(r;s_1,s_2,\ldots,s_t)$-erasure correcting if,
for any $0\leq i_1<i_2<\ldots <i_t\leq m-1$, $\C$ can correct up
to $s_j+r$ erasures in each row $i_j$ of an array in $\C$. 
We say that $\C$ is an $(r;s)$ partial-MDS (PMDS) code if, 
{\em for every} $(s_1,s_2,\ldots,s_t)$ such that each $s_j\geq 1$ 
and $\sum_{j=1}^ts_j\eq s$, %and for any $0\leq i_1<i_2<\ldots <i_t\leq m-1$, $\C$ can correct up
%to $s_j+r$ erasures in each stripe $i_j$ of an array in $\C$.
$\C$ is an
$(r;s_1,s_2,\ldots,s_t)$-erasure correcting code. 
}
\end{defin}

In the next section we give a general
construction of codes by providing their $(mr+s)\times mn$
parity-check matrices. Some of these codes are going
to be PMDS. In particular, we will analize the case $r\eq 1$
in Section~\ref{special} due to its important practical value, since
it extends RAID~5. 

\section{Code Construction}
\label{construction}

As stated in Section~\ref{PMDS}, our entries consist of $b$ bits. We
will assume that each entry is in a ring. The ring is
defined by a polynomial $f(x)$ of
degree $b$, i.e., the product of two elements in the ring (taken as
polynomials of degree up to $b-1$), is the remainder of dividing the
product of both elements by $f(x)$ (if $f(x)$ is irreducible, the
ring becomes the field $GF(2^b)$~\cite{ms}).

Let $\al$ be a root of the polynomial $f(x)$ defining the ring.
We call the {\em exponent} of $f(x)$, denoted $e(f(x))$, the exponent
of $\al$, i.e., the minimum $\ell$, $0<\ell$, such that
$\al^{\ell}\eq 1$. If $f(x)$ is primitive~\cite{ms}, $e(f(x))\eq
2^b-1$.

A special case that will be important in applications
is $f(x)\eq M_p(x)\eq 1+x+\cdots +x^{p-1}$, $p$ a prime number. In this
case, $e(M_p(x))\eq p$ and
$f(x)$ may not be irreducible. In fact, it is
not difficult to prove that $f(x)$
is irreducible if and only if 2 is primitive in $GF(p)$. So, 
the polynomials of degree up to $p-2$ modulo $M_p(x)$
constitute a ring and not generally a field.
This ring was used in~\cite{br} to construct
the Blaum-Roth (BR) codes, and for the rest of the paper, we either
assume that 
$f(x)$ is irreducible or that 
$f(x)\eq M_p(x)$, and $p$ will always denote a prime number.

We present next a general construction, and then we illustrate
it with some examples.

\begin{constr}
\label{c2}
{\em
Consider the binary polynomials modulo $f(x)$, where either $f(x)$ is
irreducible or $f(x)\eq M_p(x)$, and
let $mn\leq e(f(x))$, where $e(f(x))$ is the exponent of
$f(x)$. Let $\C(m,n,r,s;f(x))$ be the code 
whose $(mr+s)\times mn$ parity-check matrix is

{\small
\begin{eqnarray}
\label{H2}
\cH(m,n,r,s)&=&\left(
\begin{array}
{c|c|c|c}
%{c|c|c|c|c}
H(n,r,0,0)&\uzero(n,r)&%\uzero(n,r)&
\ldots &\uzero(n,r)\\
\uzero(n,r)&H(n,r,0,r)&%\uzero(n,r)&
\ldots &\uzero(n,r)\\
\vdots & \vdots &% \vdots&
\ddots &\vdots\\
\uzero(n,r)&\uzero(n,r)&
\ldots &H(n,r,0,(m-1)r)\\
\hline
\multicolumn{4}{c}{H(mn,s,r,0)}\\
\end{array}
\right)
\end{eqnarray}
}
where, if $f(\al)\eq 0$, $H(n,r,i,j)$ is the $r\times n$ matrix

\begin{eqnarray}
\label{hs2}
H(n,r,i,j)&=&\left(
\begin{array}{c|c|c|c|c}
\al^{j2^i}&\al^{(j+1)2^i}&\al^{(j+2)2^i}&\ldots &\al^{(j+n-1)2^i}\\
\al^{j2^{i+1}}&\al^{(j+1)2^{i+1}}&\al^{(j+2)2^{i+1}}&\ldots &\al^{(j+n-1)2^{i+1}}\\
\al^{j2^{i+2}}&\al^{(j+1)2^{i+2}}&\al^{(j+2)2^{i+2}}&\ldots &\al^{(j+n-1)2^{i+2}}\\
\vdots &\vdots &\vdots &\ddots &\vdots \\
\al^{j2^{i+r-1}}&\al^{(j+1)2^{i+r-1}}&\al^{(j+2)2^{i+r-1}}&\ldots &\al^{(j+n-1)2^{i+r-1}}\\
\end{array}
\right)
\end{eqnarray}
}
\end{constr}

Let us point out that matrices $H(n,r,i,j)$ as given
by~(\ref{hs2}), in which each row is the square of the previous one,
were used in~\cite{d}\cite{ga}\cite{r} for constructing codes for
which the metric is given by the rank, in~\cite{bm} for constructing
codes that can be encoded on columns and decoded on rows, and
in~\cite{lmstoa} for constructing the so called differential MDS codes.

Let us illustrate
Construction~\ref{c2} in the next example.

\begin{ex}
\label{ex2}
{\em
Consider $m\eq 3$ and $n\eq 5$, then,

{\small

\begin{eqnarray*}
\cH(3,5,1,3)&=&\left(
\begin{array}{ccccc|ccccc|ccccc}
1&1&1&1&1&0&0&0&0&0&0&0&0&0&0\\
0&0&0&0&0&1&1&1&1&1&0&0&0&0&0\\
0&0&0&0&0&0&0&0&0&0&1&1&1&1&1\\
\hline
1&\al&\al^2&\al^3&\al^4&\al^5&\al^6&\al^7&\al^8&\al^9&\al^{10}&\al^{11}&
\al^{12}&\al^{13}&\al^{14}\\
1&\al^2&\al^4&\al^6&\al^8&\al^{10}&\al^{12}&\al^{14}&\al^{16}&\al^{18}&\al^{20}&\al^{22}&
\al^{24}&\al^{26}&\al^{28}\\
1&\al^4&\al^8&\al^{12}&\al^{16}&\al^{20}&\al^{24}&\al^{28}&\al^{32}&\al^{36}&\al^{40}&\al^{44}&
\al^{48}&\al^{52}&\al^{56}\\
\end{array}
\right)\\
\end{eqnarray*}
\begin{eqnarray*}
\cH(3,5,2,2)&=&\left(
\begin{array}{ccccc|ccccc|ccccc}
1&1&1&1&1&0&0&0&0&0&0&0&0&0&0\\
1&\al&\al^2&\al^3&\al^4&0&0&0&0&0&0&0&0&0&0\\
0&0&0&0&0&1&1&1&1&1&0&0&0&0&0\\
0&0&0&0&0&\al^5&\al^6&\al^7&\al^8&\al^9&0&0&0&0&0\\
0&0&0&0&0&0&0&0&0&0&1&1&1&1&1\\
0&0&0&0&0&0&0&0&0&0&\al^{10}&\al^{11}&\al^{12}&\al^{13}&\al^{14}\\
\hline
1&\al^2&\al^4&\al^6&\al^8&\al^{10}&\al^{12}&\al^{14}&\al^{16}&\al^{18}&\al^{20}&\al^{22}&
\al^{24}&\al^{26}&\al^{28}\\
1&\al^4&\al^8&\al^{12}&\al^{16}&\al^{20}&\al^{24}&\al^{28}&\al^{32}&\al^{36}&\al^{40}&\al^{44}&
\al^{48}&\al^{52}&\al^{56}\\
\end{array}
\right)\\
\end{eqnarray*}
}
}
\end{ex}

%From now on, we denote a code $\C(m,n,r,s;f(x))$ simply as
%$\C(m,n,r,s)$ when there is no confusion regarding the polynomial
%$f(x)$; if a code $\C(m,n,r,s)$
%is $(r,s)$ PMDS, we simply say that it is PMDS.
So far we have not proved that Construction~\ref{c2} provides PMDS
codes. Actually, this is not true in 
general. The
answer depends on the particular 
parameters and on the polynomial $f(x)$ defining the ring or field.
%This fact will be illustrated in the next section, where we examine
%the case most relevant to applications, that is, $r\eq 1$.

We denote by
$(a_{i,j})_{0\leq i\leq m-1\atop 0\leq j\leq n-1}$ the received
entries from a stored array in $\C(m,n,r,s;f(x))$, assuming that the erased
ones are equal to 0. 
The first step to retrieve the erased entries consists of computing the
$rm+s$ syndromes. Using
the parity-check matrix $\cH(m,n,r,s;f(x))$ given by~(\ref{H2}), the
syndromes are

\begin{eqnarray}
\label{S0}
S_{ir}&=&\bigoplus_{j=0}^{n-1}\,a_{i,j}\;\;{\rm
for}\;\;0\leq i\leq m-1\\
\label{Si2}
S_{ir+l+1}&=&\bigoplus_{j=0}^{n-1}\,\al^{(ni+j)2^l}a_{i,j}\;\;{\rm
for}\;\;0\leq i\leq m-1\,\,,\,\,0\leq l\leq r-2\\
\label{Sm2}
S_{mr+u}&=&\bigoplus_{i=0}^{m-1}\bigoplus_{j=0}^{n-1}\,\al^{(ni+j)(2^{r+u-1})}a_{i,j}
\;\;{\rm for}\;\;0\leq u\leq s-1
\end{eqnarray}

After computing the syndromes, the erasures are recovered by solving
a linear system based on the parity-check matrix, provided that such
a solution exists. 
In the next section, we study the case $r\eq 1$ and give necessary
and sufficient conditions
that determine whether a $\C(m,n,1,s;f(x))$
code is PMDS. 

\section{The case $r\eq 1$}
\label{special}

In this section, we assume that $r\eq 1$, thus, the parity-check
matrix $\cH(m,n,1,s)$ given by~(\ref{H2}) can be written as
\begin{eqnarray*}\cH(m,n,1,s)&=&\left(
\cH_0(m,n,1,s),\cH_1(m,n,1,s),\ldots,\cH_{m-1}(m,n,1,s)
\right),\end{eqnarray*}
where, for $0\leq j\leq m-1$
\begin{eqnarray*}
\cH_j(m,n,1,s)&=&\left(
\begin{array}{cccr}
\begin{array}{r}
0\\
0\\
\vdots \\
0\\
j\ra 1\\
0\\
\vdots\\
0\\
\end{array}&
\begin{array}{c}
0\\
0\\
\vdots \\
0\\
1\\
0\\
\vdots\\
0\\
\end{array}&
\begin{array}{c}
\ldots\\
\ldots\\
\ddots \\
\ldots\\
\ldots\\
\ldots\\
\ddots\\
\ldots\\
\end{array}&
\left.
\begin{array}{c}
0\\
0\\
\vdots \\
0\\
1\\
0\\
\vdots\\
0\\
\end{array}\right\}m\\
\hspace{.8cm}\al^{jn}&\al^{jn+1}&\ldots &\al^{(j+1)n-1}\\
\hspace{.8cm}\al^{2jn}&\al^{2(jn+1)}&\ldots &\al^{2((j+1)n-1)}\\
\hspace{.8cm}\al^{4jn}&\al^{4(jn+1)}&\ldots &\al^{4((j+1)n-1)}\\
\hspace{.8cm}\vdots &\vdots &\ddots &\vdots\hspace{1.3cm} \\
\hspace{.8cm}\al^{2^{s-1}jn}&\al^{2^{s-1}(jn+1)}&\ldots &\al^{2^{s-1}((j+1)n-1)}\\
\end{array}
\right).
\end{eqnarray*}

Assume that $\sum_{j=1}^t\,s_j\eq s$ for integers $s_j\geq 1$. According to
Definition~\ref{defPMDS}, we will characterize when
$\C(m,n,1,s;f(x))$ is %PMDS.
$(r;s_1,s_2,\ldots,s_t)$-erasure correcting. 
We need a series of lemmas first.

%The following results are given without proof.

\begin{lemma}
\label{lmain0}
{\em
For $s\geq 1$, code $\C(m,n,1,s;f(x))$ as given by
Construction~\ref{c2} is PMDS if and only if:
\begin{enumerate}
\item code $\C(m,n,1,s-1;f(x))$ is PMDS, and; 
\item for any $(s_1,s_2,\ldots,s_t)$ such that 
$\sum_{j=1}^t\,s_j\eq s$, 
for any
$0\leq i_2<i_3<\ldots <i_t\leq m-1$,  
and for any  $1\leq j\leq t$ and $0\leq
l_{j,0}<l_{j,1}<\cdots <l_{j,s_j}\leq n-1$, 
\begin{eqnarray}
\label{eqmain0}
\gcd\left(\left(\sum_{u=1}^{s_1}\,\left(1+ x^{l_{1,u}-l_{1,0}}\right)\right)\,+\,
\left(\sum_{j=2}^t
x^{i_jn+l_{j,0}-l_{1,0}}\sum_{u=1}^{s_j}\,\left(1+
x^{l_{j,u}-l_{j,0}}\right)\right)\,,\,f(x)\right)\eq 1 
\end{eqnarray}

\end{enumerate}
}
\end{lemma}

\pf
Since $f(x)$ is implicit, let us denote $\C(m,n,1,s;f(x))$ simply by $\C(m,n,1,s)$.
Consider rows $0\leq i_1<i_2<\cdots <i_t\leq m-1$ such 
that row $i_j$ has exactly $s_j+1$ erasures 
in locations $(i_j,l_{j,0}),(i_j,l_{j,1}),\ldots,(i_j,l_{j,s_j}),$ for $0\leq
l_{j,0}<l_{j,1}<\cdots <l_{j,s_j}\leq n-1$ and
$\sum_{j=1}^t\,s_j\eq s$. 
%We may assume without loss of generality
%that rows $i$ with $i\neq i_j$ for $1\leq j\leq t$ have no erasures,
%since those cases can have at most one erasure, which are corrected
%by using single parity. Also without loss of generality, we may
%assume that rows $i_j$, $1\leq j\leq t$, have at least two erasures
%each, i.e., $s_j\geq 1$ for $1\leq j\leq t$. As stated above, the
Assuming the erased entries to be equal to zero and computing the
syndromes according 
to~(\ref{S0}) and~(\ref{Sm2}), we obtain

\begin{eqnarray}
\label{syn0}
\bigoplus_{v=0}^{s_j}\,a_{i_j,l_{j,v}}&=&S_{i_j}\;\;{\rm for}\;\;1\leq
j\leq t\\
\label{syn1}
\bigoplus_{j=1}^{t}\bigoplus_{v=0}^{s_j}\,\al^{2^{u}(i_jn+l_{j,v})}a_{i_j,l_{j,v}}&=&S_{m+u}\;\;{\rm
for}\;\; 0\leq u\leq s-1
\end{eqnarray}

The system given by~(\ref{syn0}) and~(\ref{syn1}) has a unique
solution if and only if the $(t+s)\times (t+s)$ matrix

\begin{eqnarray*}
\uc&=&\left(\begin{array}{c|c|c|c}
\uc_1&\uc_2&\ldots &\uc_t\\
\end{array}\right)
\end{eqnarray*}
is invertible, where $\uc_j$ is the $(t+s)\times (s_j+1)$ matrix

\begin{eqnarray*}
%\label{pcr1}
%\cH(m,n,1,s)&=&
\uc_j&=&
\left(
\begin{array}{cccc}
0&0&\ldots &0\\
0&0&\ldots &0\\
\vdots&\vdots&\ddots&\vdots\\
0&0&\ldots &0\\
j\ra 1\phantom{00}\hspace{.2cm}&1&\ldots &1\\
0&0&\ldots &0\\
0&0&\ldots &0\\
\vdots&\vdots&\ddots&\vdots\\
0&0&\ldots &0\\
\hline
\al^{i_jn+l_{j,0}}&\al^{i_jn+l_{j,1}}&\ldots &\al^{i_jn+l_{j,s_j}}\\
\al^{2(i_jn+l_{j,0})}&\al^{2(i_jn+l_{j,1})}&\ldots &\al^{2(i_jn+l_{j,s_j})}\\ 
\al^{4(i_jn+l_{j,0})}&\al^{4(i_jn+l_{j,1})}&\ldots &\al^{4(i_jn+l_{j,s_j})}\\
\vdots&\vdots&\ddots&\vdots\\
\al^{2^{s-1}(i_jn+l_{j,0})}&\al^{2^{s-1}(i_jn+l_{j,1})}&\ldots
&\al^{2^{s-1}(i_jn+l_{j,s_j})}\\  
\end{array}
\right)
\end{eqnarray*}

By row operations on $\uc$, we obtain a new $(t+s)\times (t+s)$ matrix

\begin{eqnarray*}
\uc'&=&
\left(\begin{array}{c|c|c|c}
\uc'_1&\uc'_2&\ldots &\uc'_t\\
\end{array}\right)
\end{eqnarray*}
where $\uc'_j$ is the $(t+s)\times (s_j+1)$ matrix

\begin{eqnarray*}
%\label{pcr1}
%\cH(m,n,1,s)&=&
\uc'_j&=&
\left(
\begin{array}{rccc}
0&0&\ldots &0\\
0&0&\ldots &0\\
\vdots &\vdots&\ddots&\vdots\\
0&0&\ldots &0\\
j\ra 1&1&\ldots &1\\
0&0&\ldots &0\\
0&0&\ldots &0\\
\vdots&\vdots&\ddots&\vdots\\
0&0&\ldots &0\\
\hline
0&\al^{i_jn+l_{j,0}}\left(1\xor \al^{l_{j,1}-l_{j,0}}\right)
&\ldots 
&\al^{i_jn+l_{j,0}}\left(1\xor \al^{l_{j,s_j}-l_{j,0}}\right)\\ 
0&\al^{2(i_jn+l_{j,0})}\left(1\xor \al^{2(l_{j,1}-l_{j,0})}\right)
&\ldots 
&\al^{2(i_jn+l_{j,0})}\left(1\xor \al^{2(l_{j,s_j}-l_{j,0})}\right)\\ 
0&\al^{4(i_jn+l_{j,0})}\left(1\xor \al^{4(l_{j,1}-l_{j,0})}\right)
&\ldots &\al^{4(i_jn+l_{j,0})}\left(1\xor \al^{4(l_{j,s_j}-l_{j,0})}\right)\\ 
\vdots &\vdots&\ddots&\vdots\\
0&\al^{2^{s-1}(i_jn+l_{j,0})}\left(1\xor \al^{2^{s-1}(l_{j,1}-l_{j,0})}\right)
&\ldots &\al^{2^{s-1}(i_jn+l_{j,0})}\left(1\xor \al^{2^{s-1}(l_{j,s_j}-l_{j,0})}\right)\\ 
\end{array}
\right)
\end{eqnarray*}

Notice that $\uc$
%$\left(\begin{array}{c|c|c|c}
%\uc_1&\uc_2&\ldots &\uc_t\\
%\end{array}\right)$ 
is invertible if and only if $\uc'$
%$\left(\begin{array}{c|c|c|c}
%\uc'_1&\uc'_2&\ldots &\uc'_t\\
%\end{array}\right)$ 
is invertible, if and only if the $s\times s$
matrix $\uc''\eq\left(\begin{array}{c|c|c|c}
\uc''_1&\uc''_2&\ldots &\uc''_t\\\end{array}\right)$ is invertible, where

\begin{eqnarray*}
%\label{pcr1}
%\cH(m,n,1,s)&=&
\uc''_j&=&
\left(
\begin{array}{ccc}
\al^{i_jn+l_{j,0}}\left(1\xor \al^{l_{j,1}-l_{j,0}}\right)
&\ldots 
&\al^{i_jn+l_{j,0}}\left(1\xor \al^{l_{j,s_j}-l_{j,0}}\right)\\ 
\al^{2(i_jn+l_{j,0})}\left(1\xor \al^{2(l_{j,1}-l_{j,0})}\right)
&\ldots 
&\al^{2(i_jn+l_{j,0})}\left(1\xor \al^{2(l_{j,s_j}-l_{j,0})}\right)\\ 
\al^{4(i_jn+l_{j,0})}\left(1\xor \al^{4(l_{j,1}-l_{j,0})}\right)
&\ldots &\al^{4(i_jn+l_{j,0})}\left(1\xor \al^{4(l_{j,s_j}-l_{j,0})}\right)\\ 
\vdots&\ddots&\vdots\\
\al^{2^{s-1}(i_jn+l_{j,0})}\left(1\xor \al^{2^{s-1}(l_{j,1}-l_{j,0})}\right)
&\ldots &\al^{2^{s-1}(i_jn+l_{j,0})}\left(1\xor \al^{2^{s-1}(l_{j,s_j}-l_{j,0})}\right)\\ 
\end{array}
\right)
\end{eqnarray*}

Dividing each row $u$, $0\leq u\leq s$, by $\al^{2^{u}(i_1n+l_{1,0})}$, we
obtain the $s\times s$ matrix $$\hat{\uc}\eq \left(\begin{array}{c|c|c|c}
\hat{\uc_1}&\hat{\uc_2}&\ldots &\hat{\uc_t}\\
\end{array}\right),$$ where
 
\begin{eqnarray*}
%\label{pcr1}
%\cH(m,n,1,s)&=&
\hat{\uc_1}&=&
\left(
\begin{array}{ccc}
1\xor \al^{l_{1,1}-l_{1,0}}
&\ldots 
&1\xor \al^{l_{1,s_1}-l_{1,0}}\\ 
1\xor \al^{2(l_{1,1}-l_{1,0})}
&\ldots 
&1\xor \al^{2(l_{1,s_1}-l_{1,0})}\\ 
1\xor \al^{4(l_{1,1}-l_{1,0})}
&\ldots &1\xor \al^{4(1_{1,s_1}-l_{1,0})}\\ 
\vdots&\ddots&\vdots\\
1\xor \al^{2^{s-1}(l_{1,1}-l_{1,0})}
&\ldots &1\xor \al^{2^{s-1}(l_{1,s_1}-l_{1,0})}\\ 
\end{array}
\right)
\end{eqnarray*}
and, for $2\leq j\leq t$, making $i_j\la i_j-i_1$, we have 

\begin{eqnarray*}
%\label{pcr1}
%\cH(m,n,1,s)&=&
\hat{\uc_j}&=&
\left(
\begin{array}{ccc}
\al^{i_jn+l_{j,0}-l_{1,0}}\left(1\xor \al^{l_{j,1}-l_{j,0}}\right)
&\ldots 
&\al^{i_jn+l_{j,0}-l_{1,0}}\left(1\xor \al^{l_{j,s_j}-l_{j,0}}\right)\\ 
\al^{2(i_jn+l_{j,0}-l_{1,0})}\left(1\xor \al^{2(l_{j,1}-l_{j,0})}\right)
&\ldots 
&\al^{2(i_jn+l_{j,0}-l_{1,0})}\left(1\xor \al^{2(l_{j,s_j}-l_{j,0})}\right)\\ 
\al^{4(i_jn+l_{j,0}-l_{1,0})}\left(1\xor \al^{4(l_{j,1}-l_{j,0})}\right)
&\ldots &\al^{4(i_jn+l_{j,0}-l_{1,0})}\left(1\xor \al^{4(l_{j,s_j}-l_{j,0})}\right)\\ 
\vdots&\ddots&\vdots\\
\al^{2^{s-1}(i_jn+l_{j,0}-l_{1,0})}\left(1\xor \al^{2^{s-1}(l_{j,1}-l_{j,0})}\right)
&\ldots &\al^{2^{s-1}(i_jn+l_{j,0}-l_{1,0})}\left(1\xor
\al^{2^{s-1}(l_{j,s_j}-l_{j,0})}\right)\\  
\end{array}
\right)
\end{eqnarray*}

Therefore, the $s\times s$ matrix $\hat{\uc}$ consists of a first row
$\uw_0$ followed by succesive 
rows $\uw_u$, where each row is the square of the previous row, i.e., $\uw_u\eq
\uw_0^{2^u}$ for $1\leq u\leq s-1$. Matrix $\hat{\uc}$ is invertible
if and only if its determinant is 
invertible. The determinant of a matrix of this type is
known~\cite{bm}: it is the product of the XOR of all possible subsets
of elements of the first row. For example, if we have a matrix
$$\left(\begin{array}{ccc}
\ga_1 &\ga_2 &\ga_3\\
\ga_1^2 &\ga_2^2 &\ga_3^2\\
\ga_1^4 &\ga_2^4 &\ga_3^4\\
\end{array}
\right),
$$
then its determinant is $\ga_1 \ga_2
\ga_3\left(\ga_1\xor\ga_2\right)\left(\ga_1\xor\ga_3\right)
\left(\ga_2\xor\ga_3\right)\left(\ga_1\xor\ga_2\xor\ga_2\right)$.
This result is proven similarly to the one of Vandermonde matrices.
For the sake of completeness, we prove it in the Appendix.

The first row of matrix $\hat{\uc}$ is given by 

\begin{eqnarray*}
\uw_0&=&(\uw_{0,1},\uw_{0,2},\ldots,\uw_{0,t}),
\end{eqnarray*}
where
\begin{eqnarray*}
\uw_{0,1}&=&\left(1\xor \al^{l_{1,1}-l_{1,0}},1\xor \al^{l_{1,2}-l_{1,0}},
\ldots ,1\xor \al^{l_{1,s_1}-l_{1,0}}\right)\\
\end{eqnarray*}
and, for $2\leq j\leq t$, 
\begin{eqnarray}
\label{eqw0j}
\uw_{0,j}&=&\left(\al^{i_jn+l_{j,0}-l_{1,0}}\left(1\xor
\al^{l_{j,1}-l_{j,0}}\right), 
\ldots, 
\al^{i_jn+l_{j,0}-l_{1,0}}\left(1\xor \al^{l_{j,s_j}-l_{j,0}}\right)\right).
\end{eqnarray}

%Thus, by the observations above, $\C(m,n,1,s)$ is
%$(s_1+1,s_2+1,\ldots,s_t+1)$-erasure correcting if and only if, for any
%integers $(s'_1,s'_2,\ldots,s'_t)$ such that $0\leq s'_j\leq s_j$ for each
%$1\leq j\leq t$ and at least for one $j$, $s'_j<s_j$, $\C(m,n,1,s)$ is
%$(s'_1+1,s'_2+1,\ldots,s'_t+1)$-erasure correcting and, 
Then, code $\C(m,n,1,s)$ is PMDS if and only if 
the determinant $\det\left(\hat{\uc}\right)$ is invertible, if and
only if the XOR of any subset of the elements of $\uw_0$ is
invertible. Since $\uw_0$ has $s$ elements, we may assume that if we
XOR a number elements smaller than $s$, the result is true by
induction, so assume that we take the XOR of all the $s$ elements in
$\uw_0$. Then, code $\C(m,n,1,s)$ is PMDS if and only if code
$\C(m,n,1,s-1)$ is PMDS and

\begin{eqnarray}
\label{main}
\left(\bigoplus_{u=1}^{s_1}\,\left(1\xor \al^{l_{1,u}-l_{1,0}}\right)\right)\,\xor\,
\left(\bigoplus_{j=2}^t
\al^{i_jn+l_{j,0}-l_{1,0}}\bigoplus_{u=1}^{s_j}\,\left(1\xor \al^{l_{j,u}-l_{j,0}}
\right)\right)
\end{eqnarray}
is invertible. But~(\ref{main}) is invertible if and only
if~(\ref{eqmain0}) holds.\qed

\begin{lemma}
\label{lmain}
{\em
Consider a code $\C(m,n,1,s;f(x))$ and let $s_j\geq 1$ for $1\leq j\leq t$ such that
$\sum_{j=1}^t\,s_j\eq s$. For each $s_j$, if $s_j$ is odd, let $s'_j\eq s_j$, while if $s_j$ is
even, let $s'_j\eq s_j-1$ and $s'\eq\sum_{j=1}^t\,s'_j$. Then,
$\C(m,n,1,s;f(x))$ is %PMDS if and only  if $\C(m,n,1,s';f(x))$ is PMDS.
$(1;s_1,s_2,\ldots,s_t)$-erasure correcting if and
only if $\C(m,n,1,s';f(x))$ is
$(1;s'_1,s'_2,\ldots,s'_t)$-erasure correcting.
}
\end{lemma}

\pf 
Consider~(\ref{main}). If $s_j$ is odd, 
\begin{eqnarray}
\label{main1}
\bigoplus_{u=1}^{s_j}\,\left(1\xor \al^{l_{j,u}-l_{j,0}}\right)&=&
1\xor \bigoplus_{u=1}^{s_j}\,\al^{l_{j,u}-l_{j,0}},
\end{eqnarray}
while if $s_j$ is even, 
\begin{eqnarray}
\nonumber
\bigoplus_{u=1}^{s_j}\,\left(1\xor \al^{l_{j,u}-l_{j,0}}\right)&=&
\bigoplus_{u=1}^{s_j}\,\al^{l_{j,u}-l_{j,0}}\\
\label{main2}
&=&\al^{l_{j,1}-l_{j,0}}\left(1\xor
\bigoplus_{u=2}^{s_j}\,\al^{l_{j,u}-l_{j,1}}\right)
\end{eqnarray}

If $s_1$ is even, making $s'_1\eq s_1-1$, according to~(\ref{main2}) 
and~(\ref{main1}), (\ref{main}) becomes 

\begin{eqnarray}
\nonumber
\al^{l_{1,1}-l_{1,0}}\left(1\xor
\bigoplus_{u=2}^{s_1}\,\al^{l_{1,u}-l_{1,1}}\right)\,\xor\,
\left(\bigoplus_{j=2}^t
\al^{i_jn+l_{j,0}-l_{1,0}}\bigoplus_{u=1}^{s_j}\,\left(1\xor \al^{l_{j,u}-l_{j,0}}
\right)\right)\eq\\
\nonumber
\al^{l_{1,1}-l_{1,0}}\left(\left(1\xor
\bigoplus_{u=2}^{s_1}\,\al^{l_{1,u}-l_{1,1}}\right)\,\xor\,
\left(\bigoplus_{j=2}^t
\al^{i_jn+l_{j,0}-l_{1,1}}\bigoplus_{u=1}^{s_j}\,\left(1\xor \al^{l_{j,u}-l_{j,0}}
\right)\right)\right)\eq\\
\label{main3}
\al^{l_{1,1}-l_{1,0}}\left(
\bigoplus_{u=2}^{s_1}\,\left(1\xor \al^{l_{1,u}-l_{1,0}}\right)
\,\xor\,
\left(\bigoplus_{j=2}^t
\al^{i_jn+l_{j,0}-l_{1,1}}\bigoplus_{u=1}^{s_j}\,\left(1\xor \al^{l_{j,u}-l_{j,0}}
\right)\right)\right)
\end{eqnarray}

Since $\al^{l_{1,1}-l_{1,0}}$ is always invertible, then,
by~(\ref{main}) and~(\ref{main3}), if $s_1$ is even,
$\C(m,n,1,s-1)$ is $(1;s'_1,s_2,\ldots,s_t)$-erasure correcting if
and only if
$\C(m,n,1,s)$ is $(1;s_1,s_2,\ldots,s_t)$-erasure
correcting. Similarly, if $2\leq v\leq t$ and $s_v$ is even,
according to~(\ref{main2}) and~(\ref{main1}), ~(\ref{main}) becomes 

\begin{eqnarray}
\nonumber
\left(\bigoplus_{u=1}^{s_1}\,\left(1\xor \al^{l_{1,u}-l_{1,0}}\right)\right)\,\xor\,
\left(\al^{i_vn+l_{v,0}-l_{1,0}}\bigoplus_{u=1}^{s_v}\,\left(1\xor
\al^{l_{v,u}-l_{v,0}}\right)\right)\,\xor\,\\ 
\nonumber
\left(\bigoplus_{j=2\atop j\neq v}^t
\al^{i_jn+l_{j,0}-l_{1,0}}\bigoplus_{u=1}^{s_j}\,\left(1\xor \al^{l_{j,u}-l_{j,0}}
\right)\right)\eq \\
\nonumber
\left(\bigoplus_{u=1}^{s_1}\,\left(1\xor \al^{l_{1,u}-l_{1,0}}\right)\right)\,\xor\,
\left(\al^{i_vn+l_{v,1}-l_{1,0}}\left(1\xor
\bigoplus_{u=2}^{s_v}\,\al^{l_{v,u}-l_{v,1}}\right)\right)\,\xor\,\\ 
\nonumber
\left(\bigoplus_{j=2\atop j\neq v}^t
\al^{i_jn+l_{j,0}-l_{1,0}}\bigoplus_{u=1}^{s_j}\,\left(1\xor \al^{l_{j,u}-l_{j,0}}
\right)\right)\eq \\
\nonumber
\left(\bigoplus_{u=1}^{s_1}\,\left(1\xor \al^{l_{1,u}-l_{1,0}}\right)\right)\,\xor\,
\left(\al^{i_vn+l_{v,1}-l_{1,0}}\bigoplus_{u=2}^{s_v}\,\left(1\xor
\al^{l_{v,u}-l_{v,1}} 
\right)\right)\,\xor\,\\ 
\label{main4}
\left(\bigoplus_{j=2\atop j\neq v}^t
\al^{i_jn+l_{j,0}-l_{1,0}}\bigoplus_{u=1}^{s_j}\,\left(1\xor \al^{l_{j,u}-l_{j,0}}
\right)\right)
\end{eqnarray}

By~(\ref{main}) and~(\ref{main4}), we may claim that if $s_v$ is even
for some $2\leq v\leq t$, then, making $s'_v\eq s_v-1$,   
$\C(m,n,1,s-1)$ is
$(1;s_1,s_2,\ldots,s_{v-1},s'_v,s_{v+1}\ldots,s_t)$-erasure correcting 
if and only if $\C(m,n,1,s)$ is $(s_1+1,s_2+1,\ldots,s_v+1,\ldots,s_t+1)$-erasure
correcting, completing the proof.\qed

\begin{lemma}
\label{lmain1}
{\em
Consider a code $\C(m,n,1,s;f(x))$ and let $s_j\geq 1$, each $s_j$ an odd
number for $1\leq j\leq t$ such that
$\sum_{j=1}^t\,s_j\eq s$. Then, $\C(m,n,1,s;f(x))$ is
$(1;s_1,s_2,\ldots,s_t)$-erasure correcting if and
only if, for any $0\leq i_2<i_3<\ldots <i_t\leq m-1$ and for any $0\leq
l_{j,0}<l_{j,1}<\ldots <l_{j,s_j}\leq n-1$ for each $1\leq j\leq t$, 

\begin{eqnarray}
\label{eqsmain}
\gcd\left(1+
\sum_{u=1}^{s_1}\,x^{l_{1,u}-l_{1,0}}+
\sum_{j=2}^t
x^{i_jn+l_{j,0}-l_{1,0}}\left(1+
\sum_{u=1}^{s_j}\,x^{l_{j,u}-l_{j,0}}\right)\,,\,f(x)
\right)&\eq &1
\end{eqnarray}
}
\end{lemma}

\pf Notice that in this case, (\ref{eqmain0}) becomes~(\ref{eqsmain}).\qed 

The combination of Lemmas~\ref{lmain0}, \ref{lmain} and~\ref{lmain1} gives the
following theorem:

\begin{theo}
\label{thmain}
{\em
For $s\geq 1$, code $\C(m,n,1,s;f(x))$ as given by
Construction~\ref{c2} is PMDS if and only if:
\begin{enumerate}
\item code $\C(m,n,1,s-1;f(x))$ is PMDS, and; 
\item for every $(s_1,s_2,\ldots,s_t)$ such that 
$\sum_{j=1}^t\,s_j\eq s$ and each $s_j$ is odd, 
for any
$0\leq i_2<i_3<\ldots <i_t\leq m-1$,  
and for any  $1\leq j\leq t$ and $0\leq
l_{j,0}<l_{j,1}<\cdots <l_{j,s_j}\leq n-1$, condition (\ref{eqsmain})
holds.    
\end{enumerate}
}
\end{theo}

Theorem~\ref{thmain} gives us conditions to check in order to
determine if a code $\C(m,n,1,s;f(x))$ as given by
Construction~\ref{c2} is PMDS, but by itself it does not
provide us with any family of PMDS codes. Consider the ring of polynomials
modulo $M_p(x)$, such that 
$mn<e(M_p(x))\eq p$. There are cases in which $M_p(x)$ is irreducible~\cite{br} and
the ring becomes a field (equivalently, 2 is primitive in $GF(p)$).
Notice that the 
polynomials in Theorem~\ref{thmain} have degree at most $mn-1<p-1\eq
\deg(f(x))$. Therefore, if $M_p(x)$ is irreducible then all such
polynomials are relatively prime with $M_p(x)$ and the code is PMDS.
Let us state this fact as a theorem, which provides a family of PMDS
codes (it is not known whether the number of irreducible polynomials
$M_p(x)$ is infinite):

\begin{theo}
\label{thmainBR}
{\em
Consider the code $\C(m,n,1,s;M_p(x))$ given by
Construction~\ref{c2} such that 
$M_p(x)$ is irreducible (or equivalently, 2 is primitive in $GF(p)$). 
Then, $\C(m,n,1,s;M_p(x))$ is PMDS.
}
\end{theo}

So far we have dealt with general values of $s$. In the next section we
examine special cases that are important in applications.

\section{Special cases}
\label{scases}
We examine each case into a separate subsection.

\subsection{The case $\C(m,n,1,1;f(x))$}
\label{cmn11}
Notice that $\C(m,n,1,1;f(x))$ is always PMDS, since by
Theorem~\ref{thmain}, we have to check if the binomials of type
$1+x^j$ for $1\leq j\leq n-1$ and $f(x)$ are relatively prime. This
is certainly the case when $f(x)$ is irreducible, but also when
$M_p(x)$ is reducible~\cite{br}. Let us state this result as a lemma:

\begin{lemma}
\label{th00}
{\em
Code $\C(m,n,1,1;f(x))$ is always PMDS.
}
\end{lemma}

\subsection{The case $\C(m,n,1,2;f(x))$}
\label{cmn12}
This case is important in applications, in particular, for arrays of
SSDs. Since \\ $\C(m,n,1,1;f(x))$ is PMDS, Theorem~\ref{thmain} gives the
following theorem for the case $s\eq 2$:

\begin{theo}
\label{th0}
%\label{l2}
{\em
Code $\C(m,n,1,2;f(x))$ is PMDS if and only if, for
any $1<i\leq m-1$, and for any  $0\leq l_{1,0}<l_{1,1}\leq n-1$,
$0\leq l_{2,0}<l_{2,1}\leq n-1$ ,  
 
\begin{eqnarray}
\label{eqmain}
\gcd\left(1+x^{l_{1,1}-l_{1,0}}+
x^{in+l_{2,0}-l_{1,0}}\left(1+
x^{l_{2,1}-l_{2,0}}\right)\,,\,f(x)
\right)&\eq &1
\end{eqnarray}
}
\end{theo}

Given the practical importance of this case, let us examine the
decoding (of which the encoding is a special case) in some detail.

Consider a PMDS code $\C(m,n,1,2;f(x))$, i.e., it satisfies the
conditions of Theorem~\ref{th0}.
Without loss of generality, assume that we 
either have 
three erasures in the same row $i_0$, or two pairs of erasures in
different rows $i_0$ and $i_1$, where $0\leq i_0<i_1\leq m-1$.
Consider first the case in which the three erasures occur in the same row $i_0$ and in
entries $j_0$, $j_1$ and $j_2$ of row $i_0$, $0\leq j_0<j_1<j_2$.
Assuming initially
that $a_{i_0,j_0}\eq a_{i_0,j_1}\eq a_{i_0,j_2}\eq 0$,
using~(\ref{S0}) and~(\ref{Sm2}) ((\ref{Si2}) is used only
for $r>1$), we compute the syndromes
$S_{i_0}$, $S_{m}$ and $S_{m+1}$.
Using the parity-check matrix $\cH(m,n,1,2)$ as given
by~(\ref{H2}), we have to solve the linear system
$$
\begin{array}{rrrrrrl}
a_{i_0,j_0}&\xor &a_{i_0,j_1}&\xor &a_{i_0,j_2}&=&S_{i_0}\\
\al^{i_0n+j_0}a_{i_0,j_0}&\xor &\al^{i_0n+j_1}a_{i_0,j_1}&\xor &
\al^{i_0n+j_2}a_{i_0,j_2}&=&S_m\\
\al^{2(i_0n+j_0)}a_{i_0,j_0}&\xor &\al^{2(i_0n+j_1)}a_{i_0,j_1}&\xor &
\al^{2(i_0n+j_2)}a_{i_0,j_2}&=&S_{m+1}\\
\end{array}
$$
%\begin{eqnarray*}
%a_{i_0,j_0}\xor a_{i_0,j_1}\xor a_{i_0,j_2}&=&S_{i_0}\\
%\al^{i_0n+j_0}a_{i_0,j_0}\xor \al^{i_0n+j_1}a_{i_0,j_1}\xor
%\al^{i_0n+j_2}a_{i_0,j_2}&=&S_m\\
%\al^{2(i_0n+j_0)}a_{i_0,j_0}\xor \al^{2(i_0n+j_1)}a_{i_0,j_1}\xor
%\al^{2(i_0n+j_2)}a_{i_0,j_2}&=&S_{m+1}\\
%\end{eqnarray*}
The solution to this system is 

\begin{eqnarray*}
a_{i_0,j_0}&=&
{\det\left(
\begin{array}{ccc}
S_{i_0}&1&1\\
S_m&\al^{i_0n+j_1}&\al^{i_0n+j_2}\\
S_{m+1}&\al^{2(i_0n+j_1)}&\al^{2(i_0n+j_2)}\\
\end{array}
\right)\over
\det\left(
\begin{array}{ccc}
1&1&1\\
\al^{i_0n+j_0}&\al^{i_0n+j_1}&\al^{i_0n+j_2}\\
\al^{2(i_0n+j_0)}&\al^{2(i_0n+j_1)}&\al^{2(i_0n+j_2)}\\
\end{array}
\right)}\\
a_{i_0,j_1}&=&
{\det\left(
\begin{array}{ccc}
1&S_{i_0}&1\\
\al^{i_0n+j_0}&S_{m}&\al^{i_0n+j_2}\\
\al^{2(i_0n+j_0)}&S_{m+1}&\al^{2(i_0n+j_2)}\\
\end{array}
\right)\over
\det\left(
\begin{array}{ccc}
1&1&1\\
\al^{i_0n+j_0}&\al^{i_0n+j_1}&\al^{i_0n+j_2}\\
\al^{2(i_0n+j_0)}&\al^{2(i_0n+j_1)}&\al^{2(i_0n+j_2)}\\
\end{array}
\right)}\\
a_{i_0,j_2}&=&
{\det\left(
\begin{array}{ccc}
1&1&S_{i_0}\\
\al^{i_0n+j_0}&\al^{i_0n+j_1}&S_{m}\\
\al^{2(i_0n+j_0)}&\al^{2(i_0n+j_1)}&S_{m+1}\\
\end{array}
\right)\over
\det\left(
\begin{array}{ccc}
1&1&1\\
\al^{i_0n+j_0}&\al^{i_0n+j_1}&\al^{i_0n+j_2}\\
\al^{2(i_0n+j_0)}&\al^{2(i_0n+j_1)}&\al^{2(i_0n+j_2)}\\
\end{array}
\right)}\\
\end{eqnarray*}

Since matrix
$$\left(
\begin{array}{ccc}
1&1&1\\
\al^{i_0n+j_0}&\al^{i_0n+j_1}&\al^{i_0n+j_2}\\
\al^{2(i_0n+j_0)}&\al^{2(i_0n+j_1)}&\al^{2(i_0n+j_2)}\\
\end{array}
\right)$$
is a Vandermonde matrix, 

\begin{eqnarray*}
{\det\left(
\begin{array}{ccc}
1&1&1\\
\al^{i_0n+j_0}&\al^{i_0n+j_1}&\al^{i_0n+j_2}\\
\al^{2(i_0n+j_0)}&\al^{2(i_0n+j_1)}&\al^{2(i_0n+j_2)}\\
\end{array}
\right)}&=&\al^{4i_0n+3j_0+j_1}\left(1\xor \al^{j_1-j_0}\right)
\left(1\xor \al^{j_2-j_0}\right)\left(1\xor \al^{j_2-j_1}\right)\\
\end{eqnarray*}

This determinant is easily inverted in a field, while in the ring of
elements modulo $M_p(x)$, the elements $1\xor \al^{j_1-j_0}$, $1\xor
\al^{j_2-j_0}$ and $1\xor \al^{j_2-j_1}$ can be efficiently inverted
(for details, see~\cite{br}).  

The encoding is a special case of the decoding. For instance, assume that we place
the two global parities in locations $(m-1,n-3)$ and $(m-1,n-2)$, as
depicted in Figure~\ref{fig1}. After computing the parities
$a_{i,n-1}$ for $0\leq i\leq m-2$ using single parity, we have to compute the parities
$a_{m-1,n-3}$, $a_{m-1,n-2}$ and $a_{m-1,n-1}$ using the method
above. In particular, the Vandermonde determinant becomes (making $i_0\eq m-1$,
$j_0\eq n-3$, $j_1\eq n-2$ and $j_2\eq n-1$) 
$\al^{4(m+n)-15}\left(1\xor \al\right)
\left(1\xor \al^{2}\right)\left(1\xor \al\right)\eq
\al^{4(m+n)-15}\left(1\xor \al^{4}\right)$. So, we have to invert
only $\al^{4(m+n)-15}\left(1\xor \al^{4}\right)$ for the encoding and
some operations may be 
precalculated, making the encoding very efficient. We omit the
details. 

We analize next the case of two pairs of erasures in
rows $i_0$ and $i_1$,
$0\leq i_0<i_1\leq m-1$, and assume that the
erased entries are  $a_{i_0,j_0}$ and $a_{i_0,j_1}$ in row $i_0$,
$0\leq j_0<j_1\leq n-1$, and
$a_{i_1,\ell_0}$ and $a_{i_1,\ell_1}$ in row $i_1$,
$0\leq \ell_0<\ell_1\leq n-1$.

Again using the parity-check matrix $\cH(m,n,1,2)$, we have to
solve the linear system of 4 equations with 4 unknowns

$$
\begin{array}{rrrrrrrrl}
a_{i_0,j_0}&\xor &a_{i_0,j_1}&&&&&=&S_{i_0}\\
&&&&a_{i_1,\ell_0}&\xor &a_{i_1,\ell_1}&=&S_{i_1}\\
\al^{i_0n+j_0}a_{i_0,j_0}&\xor &\al^{i_0n+j_1}a_{i_0,j_1}&\xor &
\al^{i_1n+\ell_0}a_{i_1,\ell_0}&\xor &\al^{i_1n+\ell_1}a_{i_1,\ell_1}&=&S_m\\
\al^{2(i_0n+j_0)}a_{i_0,j_0}&\xor &\al^{2(i_0n+j_1)}a_{i_0,j_1}&\xor &
\al^{2(i_1n+\ell_0)}a_{i_1,\ell_0}&\xor &\al^{2(i_1n+\ell_1)}a_{i_1,\ell_1}&=&S_{m+1}\\
\end{array}
$$
%\begin{eqnarray*}
%a_{i_0,j_0}\xor a_{i_0,j_1}&=&S_{i_0}\\
%a_{i_1,\ell_0}\xor a_{i_1,\ell_1}&=&S_{i_1}\\
%\al^{i_0n+j_0}a_{i_0,j_0}\xor \al^{i_0n+j_1}a_{i_0,j_1}\xor
%\al^{i_1n+\ell_0}a_{i_1,\ell_0}\xor \al^{i_1n+\ell_1}a_{i_1,\ell_1}&=&S_m\\
%\al^{2(i_0n+j_0)}a_{i_0,j_0}\xor \al^{2(i_0n+j_1)}a_{i_0,j_1}\xor
%\al^{2(i_1n+\ell_0)}a_{i_1,\ell_0}\xor \al^{2(i_1n+\ell_1)}a_{i_1,\ell_1}&=&S_{m+1}\\
%\end{eqnarray*}
where $S_{i_0}$ and $S_{i_1}$ are given by~(\ref{S0}) and
$S_m$ and $S_{m+1}$ are given
by~(\ref{Sm2}).
In order to solve this linear system, we need to invert the determinant
$$
\det\left(
\begin{array}{cccc}
1&1&0&0\\
0&0&1&1\\
\al^{i_0n+j_0}& \al^{i_0n+j_1}& \al^{i_1n+\ell_0}&
\al^{i_1n+\ell_1}\\
\al^{2(i_0n+j_0)}& \al^{2(i_0n+j_1)}& \al^{2(i_1n+\ell_0)}&
\al^{2(i_1n+\ell_1)}\\
\end{array}
\right).
$$
By row
operations, we can easily see that this determinant is equal
to the following determinant times a power of $\al$:

\begin{eqnarray*}
%\label{m1}
\det\left(
\begin{array}{cccc}
1&1&0&0\\
0&0&1&1\\
0&1\xor \al^{j_1-j_0}& 0&
\al^{(i_1-i_0)n+\ell_0-j_0}(1\xor \al^{\ell_1-\ell_0})\\
0&1\xor \al^{2(j_1-j_0)}& 0&
\al^{2((i_1-i_0)n+\ell_0-j_0)}(1\xor \al^{2(\ell_1-\ell_0)})\\
\end{array}
\right)
&=&\\
\det\left(
\begin{array}{cc}
1\xor \al^{j_1-j_0}& \al^{(i_1-i_0)n+\ell_0-j_0}(1\xor \al^{\ell_1-\ell_0})\\
1\xor \al^{2(j_1-j_0)}&
\al^{2((i_1-i_0)n+\ell_0-j_0)}(1\xor \al^{2(\ell_1-\ell_0)})\\
\end{array}
\right)&&
\end{eqnarray*}
Notice that this determinant corresponds to a $2\times 2$ Vandermonde
matrix, and it equals 
$1\xor \al^{j_1-j_0}\xor
\al^{(i_1-i_0)n+\ell_0-j_0}(1\xor \al^{\ell_1-\ell_0})$ times 
$\al^{(i_1-i_0)n+\ell_0-j_0}\left(1\xor
\al^{j_1-j_0}\right)\left(1\xor \al^{\ell_1-\ell_0}\right)$. 
We have seen that the latter is easy to invert.
Inverting $1\xor \al^{j_1-j_0}\xor
\al^{(i_1-i_0)n+\ell_0-j_0}(1\xor \al^{\ell_1-\ell_0})$, however, is
not as neat as inverting binomials $1\xor \al^j$ when the size $b$ of
the symbols is a large number.
Since $1+ x^{j_1-j_0}+
x^{(i_1-i_0)n+\ell_0-j_0}(1+ x^{\ell_1-\ell_0})$ and $f(x)$ are
relatively prime by Theorem~\ref{th0}, we can invert $1+ x^{j_1-j_0}+ 
x^{(i_1-i_0)n+\ell_0-j_0}(1+ x^{\ell_1-\ell_0})$ modulo $f(x)$ using
Euclid's algorithm. This operation may take some computational time,
but it is not done very often. When it is invoked, performance has
already been degraded due in general to a catastrophic failure. The
emphasis here is on data recovery and not on performance, since data
loss is not acceptable.

Let us now analize some concrete PMDS codes $\C(m,n,1,2;f(x))$.
Consider first finite fields $GF(2^b)$. In Table~\ref{t1}, we give the 
value $b$, the irreducible polynomial $f(x)$ (in octal notation), the
exponent $e(f(x))$, and values $m$ and $n$ for which the code
$\C(m,n,1,2;f(x))$ is PMDS according to Theorem~\ref{th0}. We have
not checked all possible 
irreducible polynomials, so we are not claiming that the values of
$m$ and $n$ are maximal in each case, but it is
certainly feasible to do so. For extensive tables of irreducible
polynomials, see~\cite{pw}. 

\begin{table}
$$
\begin{array}{cc}
\begin{array}{|c|c|c|c|c|}
\hline
b&f(x)&e(f(x)) &m&n\\
\hline
8&4\;3\;5&255&5&5\\
&5\;6\;7&85&7&5\\
&4\;3\;3&51&10&5\\
\hline
9&1\;0\;2\;1&511&20&6\\
&1\;2\;3\;1&73&10&7\\
\hline
10&3\;0\;2\;5&1023&21&6\\
  &&&15&7\\
\hline
11&6\;0\;1\;5&2047&29&6\\
&&&25&7\\
&&&22&8\\
&5\;3\;6\;1&2047&13&10\\
\hline
12&1\;5\;6\;4\;7&4095&67&6\\
&&&58&7\\
&&&50&8\\
&&&24&9\\
&&&22&10\\
\hline
\end{array}
&
\begin{array}{|c|c|c|c|c|}
\hline
b&f(x)&e(f(x)) &m&n\\
\hline
16&2\;2\;7\;2\;1\;5&13107&404&6\\
&&&346&7\\
&&&303&8\\
&&&269&9\\
&&&242&10\\
&&&164&11\\
&&&160&12\\
&&&59&16\\
&&&45&17\\
&&&53&18\\
&&&24&20\\
&&&19&22\\
&&&21&23\\
&&&18&24\\
&&&17&25\\
&&&16&26\\
\hline
\end{array}
\end{array}
$$
\caption{Some values of $b$, $f(x)$, $m$ and $n$ for which codes
$\C(m,n,1,2;f(x))$ are PMDS}
\label{t1}
\end{table}

Next, consider the case of codes $\C(m,n,1,2;M_p(x))$.
Theorem~\ref{thmainBR} solves the case in which $M_p(x)$ is irreducible, so
assume that $M_p(x)$ is not
irreducible, i.e., the ring is not a field. This ring was
considered for the BR codes~\cite{br} because it allows for efficient
correction of erasures for symbols of large size without 
using look-up tables
like in the case of finite fields. We need to
check all possible cases of Theorem~\ref{th0} for different values of
$m$ and $n$, $mn<p$.
%if $1+ x^{j_1-j_0}+ x^{(i_1-i_0)n+\ell_0-j_0}(1+
%x^{\ell_1-\ell_0})$ and $M_p(x)$ are relatively prime
%for $1\leq i_1-i_0\leq m-1$, $0\leq j_0<j_1\leq n-1$ and $0\leq
%\ell_0<\ell_1\leq n-1$, where $mn<p$.

\begin{table}
\begin{center}
\begin{tabular}{cc}
\begin{tabular}{|c|c|c|c|}
\hline
Prime & $m$ & $n$ &%$\C(m,n,1,2;M_p(x))$ 
PMDS?\\
\hline
17&4&4& YES\\
\hline
23&3&7& YES\\
&4&5& YES\\
%&5&4& YES\\
%&7&3& YES\\
\hline
31&5&6& NO\\
&6&5& NO\\
\hline
41&5&8& YES\\
&6&6& YES\\
&8&5& YES\\
\hline
43&5&8&YES \\
&6&7&YES \\
%&7&6&YES \\
%&8&5&YES \\
\hline
47&4&11&YES \\
&5&9&YES \\
%&6&7&YES \\
%&7&6&YES \\
%&9&5&YES \\
%&11&4&YES \\
\hline
71&7&10& YES\\
&8&8& YES\\
&10&7& YES\\
\hline
73&6&12& NO\\
&7&10& NO\\
&8&9& NO\\
&9&8& NO\\
%&10&7& NO\\
%&12&6& NO\\
\hline
79&6&13& YES\\
&7&11& YES\\
&8&9& YES\\
%&9&8& YES\\
%&11&7& YES\\
%&13&6& YES\\
\hline
89&8&11& NO\\
&9&9& NO\\
&11&8& YES\\
\hline
97&8&12& YES\\
&10&9& YES\\
&12&8& YES\\
\hline
103&9&11& YES\\
&10&10& YES\\
&11&9& YES\\
\hline
109&9&12& YES\\
&10&10& YES\\
&12&9& YES\\
\hline
113&10&11& YES\\
&11&10& YES\\
&12&9& YES\\
\hline
\end{tabular}
&
\begin{tabular}{|c|c|c|c|}
\hline
Prime & $m$ & $n$ &%$\C(m,n,1,2;M_p(x))$ 
PMDS?\\
\hline
127&11&11& YES\\
&13&9& YES\\
\hline
137&11&12& YES\\
&12&11& YES\\
&13&10& YES\\
&15&9& YES\\
&16&8& YES\\
\hline
151&15&10& YES\\
&16&9& YES\\
\hline
157&12&13& YES\\
&13&12& YES\\
&14&11& YES\\
&15&10& YES\\
&16&9& YES\\
\hline
167&12&13& YES\\
&13&12& YES\\
&15&11& YES\\
&16&10& YES\\
\hline
191&13&14& YES\\
&14&13& YES\\
&17&11& YES\\
\hline
193&16&12&YES \\
\hline
199&14&14&YES \\
&16&12&YES \\
\hline
223&15&14&YES \\
&17&13& YES\\
\hline
229&15&15&YES \\
&16&14& YES\\
\hline
233&15&15&YES \\
&16&14&YES \\
\hline
239&15&15&YES \\
&16&14& YES\\
\hline
241&16&15&YES \\
\hline
251&16&15& YES\\
&25&10& YES\\
\hline
257&16&16&YES \\
&32&8& YES\\
\hline
\end{tabular}
\end{tabular}
\caption{Values of $p$ such that 2 is not primitive in $GF(p)$, and
some codes $\C(m,n,1,2;M_p(x))$, $mn<p$. }
\label{t2}
\end{center}
\end{table}

The results are tabulated in Table~\ref{t2}, which gives the list of
primes between 17 and 257 for
which $M_p(x)$ is reducible (hence, 2 is not primitive in
$GF(p)$), together with some values of $m$ and $n$, and a statement
indicating whether the code is PMDS or not. For most such primes the codes are
PMDS. The only exceptions are 31, 73 and 89. The case $p\eq 89$ is
particularly interesting, since for $m\eq 8$ and $n\eq 11$ as well as
for $m\eq n\eq 9$, the codes are not PMDS. However, for $m\eq 11$ and
$n\eq 8$, the code is PMDS, which illustrates the fact that a code
being PMDS does not depend only on the polynomial $M_p(x)$ chosen, but
also on $m$ and $n$.

\subsection{The case $\C(m,n,1,3;f(x))$}
\label{cmn13}

There are two ways to obtain $s\eq 3$ as a sum of odd numbers: one is
$3$ itself, the other is $1+1+1$. Then, by Theorem~\ref{thmain}, we have

\begin{theo}
\label{th1}
{\em
Code $\C(m,n,1,3;f(x))$ is PMDS if and only if
code $\C(m,n,1,2;f(x))$ is PMDS, and, for $1\leq l_1<l_2<l_3\leq n-1$,  
\begin{eqnarray}
\label{eqmain10}
\gcd\left(1+ x^{l_1}+ x^{l_2}+ x^{l_3}\,,\,f(x)
\right)&\eq &1\
\end{eqnarray}
and, for any $1\leq i_2<i_3\leq m-1$, $0\leq l_{1,0}<l_{1,1}\leq
n-1$, $0\leq l_{2,0}<l_{2,1}\leq n-1$ and  
$0\leq l_{3,0}<l_{3,1}\leq n-1$,
\begin{eqnarray}
\nonumber
\gcd\left(1+x^{l_{1,1}-l_{1,0}}+
x^{i_2n+l_{2,0}-l_{1,0}}\left(1+
x^{l_{2,1}-l_{2,0}}\right)+\right.&&\\
\label{eq2}
\left. x^{i_3n+l_{3,0}-l_{1,0}}\left(1+
x^{l_{3,1}-l_{3,0}}\right)
\,,\,f(x)
\right)&\eq &1
\end{eqnarray}
}
\end{theo}

So, in order to check if code $\C(m,n,1,3;f(x))$ is PMDS, we 
start checking if code\\ $\C(m,n,1,2;f(x))$ is PMDS, like in the cases
tabulated in Tables~\ref{t1} and~\ref{t2}. Then we have to check if
the conditions~(\ref{eqmain10}) and~(\ref{eq2}) of Theorem~\ref{th1} are satisfied.

For instance, the codes $\C(m,n,1,2;f(x))$ in Table~\ref{t1} are 
PMDS, but condition~(\ref{eq2}) in Theorem~\ref{th1} is quite
restrictive and most of the entries do not correspond to
codes $\C(m,n,1,3;f(x))$ that are PMDS. 
In Table~\ref{t2}, however, several of the codes 
$\C(m,n,1,2;M_p(x))$ that are 
PMDS give codes $\C(m,n,1,3;M_p(x))$ that are also PMDS. 
We give the results in Table~\ref{t22}, which shows that for the
primes 17, 43, 89, 127, 151, 241 and 257, and also for 89 with
$(m,n)\eq (11,8)$, the codes $\C(m,n,1,3;M_p(x))$ are not PMDS, although
the corresponding codes $\C(m,n,1,2;M_p(x))$ were PMDS. 

\begin{table}
\begin{center}
\begin{tabular}{cc}
\begin{tabular}{|c|c|c|c|}
\hline
Prime & $m$ & $n$ &%$\C(m,n,1,3;M_p(x))$ 
PMDS?\\
\hline
17&4&4& NO\\
\hline
23&3&7& YES\\
&4&5& YES\\
%&5&4& YES\\
%&7&3& YES\\
\hline
31&5&6& NO\\
&6&5& NO\\
\hline
41&5&8& YES\\
&6&6& YES\\
&8&5& YES\\
\hline
43&5&8&NO \\
&6&7&NO \\
%&7&6&YES \\
%&8&5&YES \\
\hline
47&4&11&YES \\
&5&9&YES \\
%&6&7&YES \\
%&7&6&YES \\
%&9&5&YES \\
%&11&4&YES \\
\hline
71&7&10& YES\\
&8&8& YES\\
&10&7& YES\\
\hline
73&6&12& NO\\
&7&10& NO\\
&8&9& NO\\
&9&8& NO\\
%&10&7& NO\\
%&12&6& NO\\
\hline
79&6&13& YES\\
&7&11& YES\\
&8&9& YES\\
%&9&8& YES\\
%&11&7& YES\\
%&13&6& YES\\
\hline
89&8&11& NO\\
&9&9& NO\\
&11&8& NO\\
\hline
97&8&12& YES\\
&10&9& YES\\
&12&8& YES\\
\hline
103&9&11& YES\\
&10&10& YES\\
&11&9& YES\\
\hline
\end{tabular}
&
\begin{tabular}{|c|c|c|c|}
\hline
Prime & $m$ & $n$ &%$\C(m,n,1,3;M_p(x))$ 
PMDS?\\
\hline
109&9&12& YES\\
&10&10& YES\\
&12&9& YES\\
\hline
113&10&11&YES \\
&11&10&YES \\
&12&9&YES \\
\hline
127&11&11& NO\\
&13&9& NO\\
\hline
137&11&12&YES \\
&12&11&YES \\
%&13&10& \\
%&15&9& \\
%&16&8& \\
\hline
151&15&10&NO \\
&16&9& NO\\
\hline
157&12&13& YES\\
%&13&12& \\
%&14&11& \\
%&15&10& \\
&16&9&YES \\
\hline
%167&12&13& \\
%&13&12& \\
%&15&11& \\
167&16&10& YES\\
\hline
%191&13&14& \\
%&14&13& \\
191&17&11& YES\\
\hline
193&16&12& YES\\
\hline
%199&14&14& \\
199&16&12&YES \\
\hline
%223&15&14& \\
223&17&13&YES \\
\hline
%229&15&15& \\
229&16&14& YES\\
&28&8&YES\\
\hline
%233&15&15& \\
%&16&14& \\
233&23&10& YES\\
\hline
%239&15&15& \\
%&16&14& \\
239&26&9& YES\\
\hline
241&16&15& NO\\
&24&10& NO\\
\hline
%251&16&15& \\
251&25&10& YES\\
\hline
257&16&16&NO \\
&32&8& NO\\
\hline
\end{tabular}
\end{tabular}
\caption{%Values of $p$ such that 2 is not primitive in $GF(p)$, and
%some codes $\C(m,n,1,3;M_p(x))$, $mn<p$. 
Some codes $\C(m,n,1,3;M_p(x))$ such that $p\leq 257$ and
$M_p(x)$ is not irreducible }
\label{t22}
\end{center}
\end{table}

\subsection{The case $\C(m,n,1,4;f(x))$}
\label{c2mn14}

As done in Subsection~\ref{cmn13}, we have to start writing 
$s\eq 4$ as all possible sums of odd numbers. There are three ways of
doing so: $4\eq 1+3$, $4\eq 3+1$ and $4\eq 1+1+1+1$.  
By Theorem~\ref{thmain}, we have: 

\begin{theo}
\label{th3}
{\em
Code $\C(m,n,1,4;f(x))$ as given by
Construction~\ref{c2} is PMDS if and only if
code $\C(m,n,1,3;f(x))$ is PMDS, and, for any $1\leq i\leq m-1$, $0\leq
l_{1,0}<l_{1,1}\leq n-1$ and 
$0\leq l_{2,0}<l_{2,1}<l_{2,2}<l_{2,3}\leq n-1$, 
\begin{eqnarray*}
\label{eqmain3}
\gcd\left(1+x^{l_{1,1}-l_{1,0}}+
x^{in+l_{2,0}-l_{1,0}}\left(1+
x^{l_{2,1}-l_{2,0}}+x^{l_{2,2}-l_{2,0}}+x^{l_{2,3}-l_{2,0}}\right)
%\right)
\,,\,f(x)
\right)&\eq &1,
\end{eqnarray*}
for any $1\leq i\leq m-1$, $0\leq
l_{1,0}<l_{1,1}<l_{1,2}<l_{1,3}\leq n-1$ and 
$0\leq l_{2,0}<l_{2,1}\leq n-1$, 
\begin{eqnarray*}
\label{eqmain4}
\gcd\left(1+x^{l_{1,1}-l_{1,0}}+x^{l_{1,2}-l_{1,0}}+x^{l_{1,3}-l_{1,0}}+
x^{in+l_{2,0}-l_{1,0}}\left(1+
x^{l_{2,1}-l_{2,0}}\right)
%\right)
\,,\,f(x)
\right)&\eq &1,
\end{eqnarray*}
and for any $1\leq i_2<i_3<i_4\leq m-1$, $0\leq
l_{1,0}<l_{1,1}\leq n-1$, 
$0\leq l_{2,0}<l_{2,1}\leq n-1$, 
$0\leq l_{3,0}<l_{3,1}\leq n-1$ and $0\leq l_{4,0}<l_{4,1}\leq n-1$, 
\begin{eqnarray*}
\nonumber
\gcd\left(1+x^{l_{1,1}-l_{1,0}}+
x^{i_2n+l_{2,0}-l_{1,0}}\left(1+x^{l_{2,1}-l_{2,0}}\right)+
\right.&&\\
\label{eqmain5}
\left.
x^{i_3n+l_{3,0}-l_{1,0}}\left(1+x^{l_{3,1}-l_{3,0}}\right)+ 
x^{i_4n+l_{4,0}-l_{1,0}}\left(1+x^{l_{4,1}-l_{4,0}}\right)
%\right)
\,,\,f(x)
\right)&\eq &1.
\end{eqnarray*}
}
\end{theo}

Consider next a restricted situation for a code $\C(m,n,1,4;f(x))$.
In~\cite{lmstoa}, codes were constructed that can recover from an
erased column together with a row with up to two errors, or two
different rows with up to one error each. 
From our coding point of view, a $\C(m,n,1,4;f(x))$ code 
that is both (1;4)-erasure correcting and (1;2,2)-erasure correcting will
accomplish this (these conditions are actually stronger than those
in~\cite{lmstoa}, since they do not require an erased column, the
erasures can be anywhere in the row). For reasons of space, we don't
address at this point the decoding algorithm for errors and we
concentrate on the existence of such a code.

Notice that, according to Lemma~\ref{lmain1}, a code $\C(m,n,1,4;f(x))$ is
(1;4)-erasure correcting, if and
only if for any $1\leq l_1<l_2<l_3\leq n-1$, (\ref{eqmain10}) holds.

Also by Lemma~\ref{lmain1}, a code $\C(m,n,1,4;f(x))$ is
(1;2,2)-erasure correcting, if and
only if for any $1\leq i\leq m-1$ and $0\leq l_{1,0}<l_{1,1}\leq
n-1$, $0\leq l_{2,0}<l_{2,1}\leq n-1$, (\ref{eqmain}) holds. But
(\ref{eqmain}) is exactly the condition for code $\C(m,n,1,2)$ 
to be PMDS by Theorem~\ref{th0}. 
Thus, we have the following lemma:

\begin{lemma}
\label{lmain3}
{\em
Code $\C(m,n,1,4;f(x))$ is both (1;4)-erasure correcting
and (1;2,2)-erasure correcting if and only if
code $\C(m,n,1,2;f(x))$ is PMDS and, for any $1\leq l_1<l_2<l_3\leq n-1$,
(\ref{eqmain10}) holds.
}
\end{lemma}

We can verify in Table~\ref{t2} that for the codes
$\C(m,n,1,2;M_p(x))$ that are PMDS, (\ref{eqmain10}) holds.
Therefore, by Lemma~\ref{lmain3}, the corresponding
codes $\C(m,n,1,4;M_p(x))$ are both (1;4)-erasure correcting 
and (1;2,2)-erasure correcting.

\subsection{The case $\C(m,n,r,1;f(x))$}
\label{cmnr1}
So far, in this section we have considered cases in which $r\eq 1$.
If $r\eq s\eq 1$, we have seen in Subsection~\ref{cmn11}
that the code is PMDS, so we examine here the  case $r>1$. Thus, assume
that row $i$, $0\leq i\leq m-1$, has $r+1$ erasures in locations
$0\leq j_0<j_1<\ldots <j_r\leq n-1$. The following theorem is given
without proof (it is proven similarly to the previous cases by
examining determinants):

\begin{theo}
\label{tcmnr1}
{\em
Consider code $\C(m,n,r,1;f(x))$. If $r$ is even, then $\C(m,n,r,1;f(x))$ is
PMDS if and only if $\C(m,n,r-1,1;f(x))$ is PMDS, while if $r$ is odd, 
$\C(m,n,r,1;f(x))$ is
PMDS if and only if $\C(m,n,r-1,1;f(x))$ is PMDS and, for any $1\leq
l_1<l_2<\ldots <l_r\leq n-1$, 

\begin{eqnarray}
\label{eqcmnr1}
\gcd\left(1+\sum_{u=1}^rx^{l_u}\,,\,f(x)
\right)&\eq &1
\end{eqnarray}
}
\end{theo}

Since $\C(m,n,1,1;f(x))$ is
PMDS, by Theorem~\ref{tcmnr1}, also $\C(m,n,2,1;f(x))$ is
PMDS. According to~(\ref{eqcmnr1}), $\C(m,n,3,1;f(x))$ and $\C(m,n,4,1;f(x))$
are PMDS if and only if, for any $1\leq l_1<l_2<l_3\leq n-1$,
(\ref{eqmain10}) holds.

\subsection{The case $\C(m,n,2,2;f(x))$}
\label{cmn22}

For $\C(m,n,2,2;f(x))$ to be PMDS, it has to be both (2;2)-erasure correcting
and (2;1,1)-erasure correcting. As in the previous subsection,
$\C(m,n,2,2;f(x))$ will be (2;2)-erasure correcting if and only if, for any
$1\leq l_1<l_2<l_3\leq n-1$, (\ref{eqmain10}) holds. We have also
seen at the end of the previous subsection that this is equivalent to
saying that code $\C(m,n,3,1;f(x))$ is PMDS. By examining the
conditions under which code $\C(m,n,2,2;f(x))$ is
(2;1,1)-erasure correcting, we have the following theorem (again,
without proof):

\begin{theo}
\label{th4}
{\em
Code $\C(m,n,2,2;f(x))$ is PMDS if and only if code $\C(m,n,3,1;f(x))$ is PMDS
and, for any $1\leq i\leq m-1$, $0\leq l_{1,0}<l_{1,1}<l_{1,2}\leq n-1$ and $0\leq
l_{2,0}<l_{2,1}<l_{2,2}\leq n-1$, if 
\begin{eqnarray*}
%\nonumber
g(x)&=&1+ x^{l_{1,1}-l_{1,0}}+
x^{l_{1,2}-l_{1,0}}+x^{2(l_{1,1}-l_{1,0})}+
x^{2(l_{1,2}-l_{1,0})}+x^{(l_{1,1}-l_{1,0})+(l_{1,2}-l_{1,0})}+  
\\
&&x^{2(in+l_{2,0}-l_{1,0})}\left(1+ x^{l_{2,1}-l_{2,0}}+
%\label{eqg}
x^{l_{2,2}-l_{2,0}}+x^{2(l_{2,1}-l_{2,0})}+
x^{2(l_{2,2}-l_{2,0})}+x^{(l_{2,1}-l_{2,0})+(l_{2,2}-l_{2,0})}\right), %\\ 
\end{eqnarray*}
then $\gcd (g(x),f(x))\eq 1$.
}
\end{theo}

\section{An alternative construction}
\label{salt}
In this section we present an alternative to Construction~\ref{c2}.

\begin{constr}
\label{c1}
{\em
Consider the binary polynomials modulo $f(x)$, where either $f(x)$ is
irreducible or $f(x)\eq M_p(x)$, and
let $mn\leq e(f(x))$. Let $\C^{(1)}(m,n,r,s;f(x))$ be the code
whose $(mr+s)\times mn$ parity-check matrix is

{\small
\begin{eqnarray}
\label{H1}
\cH^{(1)}(m,n,r,s)&=&\left(
\begin{array}
{c|c|c|c}
%{c|c|c|c|c}
H^{(1)}(n,r,0,0)&\uzero(n,r)&%\uzero(n,r)&
\ldots &\uzero(n,r)\\
\uzero(n,r)&H^{(1)}(n,r,0,r)&%\uzero(n,r)&
\ldots &\uzero(n,r)\\
%\uzero(n,r)&\uzero(n,r)&H^{(1)}(n,r,0,2r)&
%\ldots &\uzero(n,r)\\
\vdots & \vdots & %\vdots&
\ddots &\vdots\\
\uzero(n,r)&\uzero(n,r)&%\uzero(n,r)&
\ldots &H^{(1)}(n,r,0,(m-1)r)\\
\hline
\multicolumn{4}{c}{H^{(1)}(mn,s,r,0)}\\
\end{array}
\right)
\end{eqnarray}
}
where, if $f(\al)\eq 0$, $H^{(1)}(n,r,i,j)$ is the $r\times n$ matrix

\begin{eqnarray}
\label{hs}
H^{(1)}(n,r,i,j)&=&\left(
\begin{array}{c|c|c|c|c}
\al^{ij}&\al^{i(j+1)}&\al^{i(j+2)}&\ldots &\al^{i(j+n-1)}\\
\al^{(i+1)j}&\al^{(i+1)(j+1)}&\al^{(i+1)(j+2)}&\ldots &\al^{(i+1)(j+n-1)}\\
\al^{(i+2)j}&\al^{(i+2)(j+1)}&\al^{(i+2)(j+2)}&\ldots &\al^{(i+2)(j+n-1)}\\
\vdots &\vdots &\vdots &\ddots &\vdots \\
\al^{(i+r-1)j}&\al^{(i+r-1)(j+1)}&\al^{(i+r-1)(j+2)}&\ldots &\al^{(i+r-1)(j+n-1)}\\
\end{array}
\right)
\end{eqnarray}
and $\uzero(n,r)$ is an $r\times n$ zero matrix.
}
\end{constr}

Next we illustrate Construction~\ref{c1} with some examples. 

\begin{ex}
\label{ex1}
{\em
Consider $m\eq 3$ and $n\eq 5$, then,

{\small
\begin{eqnarray*}
\cH^{(1)}(3,5,1,3)&=&\left(
\begin{array}{ccccc|ccccc|ccccc}
1&1&1&1&1&0&0&0&0&0&0&0&0&0&0\\
0&0&0&0&0&1&1&1&1&1&0&0&0&0&0\\
0&0&0&0&0&0&0&0&0&0&1&1&1&1&1\\
\hline
1&\al&\al^2&\al^3&\al^4&\al^5&\al^6&\al^7&\al^8&\al^9&\al^{10}&\al^{11}&
\al^{12}&\al^{13}&\al^{14}\\
1&\al^2&\al^4&\al^6&\al^8&\al^{10}&\al^{12}&\al^{14}&\al^{16}&\al^{18}&\al^{20}&\al^{22}&
\al^{24}&\al^{26}&\al^{28}\\
1&\al^3&\al^6&\al^{9}&\al^{12}&\al^{15}&\al^{18}&\al^{21}&\al^{24}&\al^{27}&\al^{30}&\al^{33}&
\al^{36}&\al^{39}&\al^{42}\\
\end{array}
\right)\\
\end{eqnarray*}
\begin{eqnarray*}
\cH^{(1)}(3,5,3,1)&=&\left(
\begin{array}{ccccc|ccccc|ccccc}
1&1&1&1&1&0&0&0&0&0&0&0&0&0&0\\
1&\al&\al^2&\al^3&\al^4&0&0&0&0&0&0&0&0&0&0\\
1&\al^2&\al^4&\al^6&\al^8&0&0&0&0&0&0&0&0&0&0\\
0&0&0&0&0&1&1&1&1&1&0&0&0&0&0\\
0&0&0&0&0&\al^5&\al^6&\al^7&\al^8&\al^9&0&0&0&0&0\\
0&0&0&0&0&\al^{10}&\al^{12}&\al^{14}&\al^{16}&\al^{18}&0&0&0&0&0\\
0&0&0&0&0&0&0&0&0&0&1&1&1&1&1\\
0&0&0&0&0&0&0&0&0&0&\al^{10}&\al^{11}&\al^{12}&\al^{13}&\al^{14}\\
0&0&0&0&0&0&0&0&0&0&\al^{20}&\al^{22}&\al^{24}&\al^{26}&\al^{28}\\
\hline
1&\al^3&\al^6&\al^9&\al^{12}&\al^{15}&\al^{18}&\al^{21}&\al^{24}&\al^{27}&\al^{30}&\al^{33}&
\al^{36}&\al^{39}&\al^{42}\\
\end{array}
\right)\\
\end{eqnarray*}
\begin{eqnarray*}
\cH^{(1)}(3,5,2,2)&=&\left(
\begin{array}{ccccc|ccccc|ccccc}
1&1&1&1&1&0&0&0&0&0&0&0&0&0&0\\
1&\al&\al^2&\al^3&\al^4&0&0&0&0&0&0&0&0&0&0\\
0&0&0&0&0&1&1&1&1&1&0&0&0&0&0\\
0&0&0&0&0&\al^5&\al^6&\al^7&\al^8&\al^9&0&0&0&0&0\\
0&0&0&0&0&0&0&0&0&0&1&1&1&1&1\\
0&0&0&0&0&0&0&0&0&0&\al^{10}&\al^{11}&\al^{12}&\al^{13}&\al^{14}\\
\hline
1&\al^2&\al^4&\al^6&\al^8&\al^{10}&\al^{12}&\al^{14}&\al^{16}&\al^{18}&\al^{20}&\al^{22}&
\al^{24}&\al^{26}&\al^{28}\\
1&\al^3&\al^6&\al^9&\al^{12}&\al^{15}&\al^{18}&\al^{21}&\al^{24}&\al^{27}&\al^{30}&\al^{33}&
\al^{36}&\al^{39}&\al^{42}\\
\end{array}
\right)\\
\end{eqnarray*}
}
}
\end{ex}

Notice that $\C(m,n,1,2;f(x))$ and $\C^{(1)}(m,n,1,2;f(x))$ coincide. Let us
analize in the next subsections some special cases.

\subsection{The case $\C^{(1)}(m,n,r,1;f(x))$}
\label{c1mnr1}
Like in Subsection~\ref{cmnr1}, we have to examine under which
conditions code $\C^{(1)}(m,n,r,1;f(x))$ is $(1;r)$-erasure correcting.
Using the parity-check matrix
$\cH^{(1)}(m,n,r,1)$ as defined by~(\ref{H1}), $\C^{(1)}(m,n,r,1;f(x))$ is
$(r;1)$-erasure correcting if and only if, for any $0\leq i\leq m-1$
and for any $1\leq j_0<j_1<\ldots <j_r$, the Vandermonde determinant

\begin{eqnarray*}
\det
\left(
\begin{array}{cccc}
1&1&\ldots &1\\
\al^{in+j_0}&\al^{in+j_1}&\ldots &\al^{in+j_r}\\
\al^{2(in+j_0)}&\al^{2(in+j_1)}&\ldots &\al^{2(in+j_r)}\\
\al^{3(in+j_0)}&\al^{3(in+j_1)}&\ldots &\al^{3(in+j_r)}\\
\vdots &\vdots &\ddots &\vdots \\
\al^{r(in+j_0)}&\al^{r(in+j_1)}&\ldots &\al^{r(in+j_r)}\\
\end{array}
\right)&\eq &\prod_{0\leq t<l\leq r}\,\left(\al^{in+j_t}\xor \al^{in+j_l}\right)
\end{eqnarray*}
is invertible. Since this is always the case, we have the
following theorem: 

\begin{theo}
\label{tc1mnr1}
{\em
Code $\C^{(1)}(m,n,r,1;f(x))$ is PMDS.
}
\end{theo}

Comparing Theorems~\ref{tcmnr1} and~\ref{tc1mnr1}, we conclude that
codes $\C^{(1)}(m,n,r,1;f(x))$ are preferable to codes $\C(m,n,r,1;f(x))$ for
$r\geq 2$, since the former are PMDS without restrictions. 

\subsection{The case $\C^{(1)}(m,n,1,3;f(x))$}
\label{c1mn13}

We give the following theorem without proof:

\begin{theo}
\label{th01}
{\em
Code $\C^{(1)}(m,n,1,3;f(x))$ as given by
Construction~\ref{c1} is PMDS if and only if, for
any $0\leq i_1\neq i_2\leq m-1$, $0\leq l_{1,0}<l_{1,1}<l_{1,2}\leq n-1$ and $0\leq
l_{2,0}<l_{2,1}\leq n-1$, $f(\al)\eq 0$, the following matrix %~(\ref{m2}) 
is invertible, 
\begin{equation}
\label{m2}
\left(
\begin{array}{lll}
1\xor \al^{l_{1,1}-l_{1,0}}&1\xor \al^{l_{1,2}-l_{1,0}}&
\al^{(i_2-i_1)n+l_{2,0}-l_{1,0}}(1\xor \al^{l_{2,1}-l_{2,0}})\\
1\xor \al^{2(l_{1,1}-l_{1,0})}&1\xor \al^{2(l_{1,2}-l_{1,0})}&
\al^{2((i_2-i_1)n+l_{2,0}-l_{1,0})}(1\xor \al^{2(l_{2,1}-l_{2,0})})\\
1\xor \al^{3(l_{1,1}-l_{1,0})}&1\xor \al^{3(l_{1,2}-l_{1,0})}&
\al^{3((i_2-i_1)n+l_{2,0}-l_{1,0})}(1\xor \al^{3(l_{2,1}-l_{2,0})})\\
\end{array}
\right)
\end{equation}
and for any 
$1\leq i_2<i_3\leq m-1$, $0\leq l_{1,0}<l_{1,1}\leq n-1$, $0\leq
l_{2,0}<l_{2,1}\leq n-1$ and $0\leq
l_{3,0}<l_{3,1}\leq n-1$,
the following matrix %~(\ref{m133}) 
is invertible: 
\begin{equation}
\label{m133}
\left(
\begin{array}{ccc}
1\xor \al^{l_{1,1}-l_{1,0}}&\al^{i_2n+l_{2,0}-l_{1,0}}(1\xor
\al^{l_{2,1}-l_{2,0}}) &\al^{i_3n+l_{3,0}-l_{1,0}}(1\xor \al^{l_{3,1}-l_{3,0}})\\
1\xor \al^{2(l_{1,1}-l_{1,0})}
&\al^{2(i_2n+l_{2,0}-l_{1,0})}(1\xor \al^{2(l_{2,1}-l_{2,0})})
&\al^{2(i_3n+l_{3,0}-l_{1,0})}(1\xor \al^{2(l_{3,1}-l_{3,0})})\\
1\xor \al^{3(l_{1,1}-l_{1,0})}
&\al^{3(i_2n+l_{2,0}-l_{1,0})}(1\xor \al^{3(l_{2,1}-l_{2,0})})
&\al^{3(i_3n+l_{3,0}-l_{1,0})}(1\xor \al^{3(l_{3,1}-l_{3,0})})\\
\end{array}
\right)
\end{equation}
}
\end{theo}

Consider
$f(x)\eq M_p(x)$. We tested all
prime numbers $p$ such that $2$ is primitive in 
$GF(p)$ up to $p\eq 227$ (i.e., $f(x)$ is irreducible), and we
found out that the matrices given by~(\ref{m2}) and~(\ref{m133}) are
invertible in all instances. Thus, we have the following lemma:

\begin{lemma}
\label{l102}
{\em
Consider the code $\C^{(1)}(m,n,1,3;M_p(x)$ given by
Construction~\ref{c1} such that
$M_p(x)$ is irreducible (or equivalently, 2 is primitive in $GF(p)$). 
Then, for $19\leq p\leq 227$, code $\C^{(1)}(m,n,1,3;M_p(x))$ is PMDS.
}
\end{lemma}

We leave as an open problem whether codes $\C^{(1)}(m,n,1,3;M_p(x))$ are
PMDS when $M_p(x)$ is 
irreducible (this result was true for codes $\C(m,n,1,3;M_p(x))$ by
Theorem~\ref{thmainBR}). 
For values of $p$ such that 2 is not primitive in $GF(p)$, some
results are tabulated 
in Table~\ref{t12} for different values of $m$ and $n$. This table is
very similar to Table~\ref{t22}. 

\begin{table}
\begin{center}
\begin{tabular}{cc}
\begin{tabular}{|c|c|c|c|}
\hline
Prime & $m$ & $n$ &%$\C^{(1)}(m,n,1,3;f(x))$ 
PMDS?\\
\hline
17&4&4& NO\\
\hline
23&3&7& NO\\
&4&5& YES\\
%&5&4& YES\\
%&7&3& YES\\
\hline
31&5&6& NO\\
&6&5& NO\\
\hline
41&5&8& NO\\
&6&6& YES\\
&8&5& YES\\
\hline
43&5&8&NO \\
&6&7&NO \\
%&7&6&YES \\
%&8&5&YES \\
\hline
47&4&11&YES \\
&5&9& YES\\
%&6&7&YES \\
%&7&6&YES \\
%&9&5&YES \\
%&11&4&YES \\
\hline
71&7&10& YES\\
&8&8& YES\\
&10&7& YES\\
\hline
73&6&12& NO\\
&7&10& NO\\
&8&9& NO\\
&9&8& NO\\
\hline
79&6&13& YES\\
&7&11&YES \\
&8&9&YES \\
\hline
89&8&11& NO\\
&9&9& NO\\
&11&8& NO\\
\hline
97&8&12&YES \\
&10&9& YES\\
&12&8& YES\\
\hline
103&9&11&YES \\
&10&10& YES\\
&11&9&YES \\
\hline
\end{tabular}
&
\begin{tabular}{|c|c|c|c|}
\hline
Prime & $m$ & $n$ &%$\C^{(1)}(m,n,1,3;f(x))$ 
PMDS?\\
\hline
109&9&12&YES \\
&10&10&YES \\
&12&9& YES\\
\hline
113&10&11& NO\\
&11&10& NO\\
&12&9& NO\\
\hline
127&11&11& NO\\
&13&9& NO\\
\hline
137&11&12&YES\\
&12&11&YES \\
&13&10& YES\\
&15&9& YES\\
&16&8& YES\\
\hline
151&15&10&NO \\
&16&9& NO\\
\hline
157&12&13& YES\\
&13&12& YES\\
%&14&11& \\
%&15&10& \\
&16&9&YES \\
\hline
%167&16&13& \\
%&13&12& \\
%&15&11& \\
167&16&10& YES\\
\hline
%191&13&14& \\
%&14&13& \\
191&17&11& YES\\
\hline
193&16&12& YES\\
\hline
%199&14&14& \\
199&16&12& YES\\
\hline
%223&15&14& \\
223&17&13& YES\\
\hline
229&16&14&YES \\
&28&8& YES\\
\hline
%233&15&15& \\
%&16&14& \\
233&23&10&YES\\
\hline
%239&15&15& \\
239&26&9& YES\\
\hline
%241&16&15& \\
241&24&10& NO  \\
\hline
%251&16&15& \\
251&25&10& YES\\
\hline
257&16&16&NO \\
&32&8& NO\\
\hline
\end{tabular}
\end{tabular}
\caption{%Values of $p\leq 257$ for which $M_p(x)$ is not irreducible 
%and some codes $\C^{(1)}(m,n,1,3;M_p(x))$, $mn<p$. 
Some codes $\C^{(1)}(m,n,1,3;M_p(x))$ such that $p\leq 257$ and
$M_p(x)$ is not irreducible 
}
\label{t12}
\end{center}
\end{table}

Comparing Tables~\ref{t22} and~\ref{t12}, we can see that for values of $p$,
$m$ and $n$ for which \\
$\C^{(1)}(m,n,1,3;M_p(x))$ is PMDS, also $\C(m,n,1,3;M_p(x))$
is PMDS. However, for $p\eq 23$, $\C(3,7,1,3;M_p(x))$ is PMDS but
$\C^{(1)}(3,7,1,3;M_p(x))$ is not, for $p\eq 41$, $\C(5,8,1,3;M_p(x))$ is PMDS but
$\C^{(1)}(5,8,1,3;M_p(x))$ is not, and for $p\eq 113$, $\C(3,7,10,11;M_p(x))$,\\
$\C(3,7,11,10;M_p(x))$ and $\C(3,7,12,9;M_p(x))$ are PMDS but
$\C^{(1)}(3,7,10,11;M_p(x))$,\\
$\C^{(1)}(3,7,11,10;M_p(x))$ and $\C^{(1)}(3,7,12,9;M_p(x))$ are not.

\section{A Simplified Construction}
\label{simple}
In this section we present a construction that is an alternative to
codes $\C(m,n,1,s;f(x))$ for $1\leq s\leq 2$. In the case of $s\eq
2$, the new construction can correct the situation depicted at the
left of Figure~\ref{fig2}, that is, two pairs of erasures in two
different rows. It cannot correct the situation at the right of 
Figure~\ref{fig2}, i.e., three erasures in the same row. This is a
tradeoff, since the new construction, as we will see, uses a smaller
finite field or ring. Explicitly:

\begin{constr}
\label{c3}
{\em
Consider the binary polynomials modulo $f(x)$, where either $f(x)$ is
irreducible or $f(x)\eq M_p(x)$, and
let $\max\{m,n\}\leq e(f(x))$, where $e(f(x))$ is the exponent of
$f(x)$. Let $\C^{(2)}(m,n,1,2;f(x))$ be the code 
whose $(m+2)\times mn$ parity-check matrix is

\begin{eqnarray*}
%\label{H3}
\cH^{(2)}(m,n,1,2)&=&
\left(
{\footnotesize %\scriptsize
\begin{array}{cccc|cccc|c|cccc}
1&1&\ldots &1&0&0&\ldots &0&\ldots &0&0&\ldots &0\\
0&0&\ldots &0&1&1&\ldots &1&\ldots &0&0&\ldots &0\\
\vdots &\vdots &\vdots &\vdots &
\vdots &\vdots &\vdots &\vdots &\ddots &
\vdots &\vdots &\vdots &\vdots \\
0&0&\ldots &0&0&0&\ldots &0&\ldots &1&1&\ldots &1\\
\hline
1&\al &\ldots &\al^{n-1}&1&\al &\ldots &\al^{n-1}&\ldots &
1&\al &\ldots &\al^{n-1}\\
1&\al &\ldots &\al^{n-1}&\al&\al^2 &\ldots &\al^{n}&\ldots &
\al^{m-1}&\al^{m-2} &\ldots &\al^{m+n-2}\\
\end{array}}
\right)
\end{eqnarray*}
$\C^{(2)}(m,n,1,1;f(x))$ is the code 
whose $(m+1)\times mn$ parity-check matrix is given by the first
$m+1$ rows of $\cH^{(2)}(m,n,1,2)$.
}
\end{constr}
 
The following example illustrates Construction~\ref{c3}.

\begin{ex}
\label{exs}
{\em
Consider codes $\C^{(2)}(3,5,1,2;M_5(x))$ and
$\C^{(2)}(5,3,1,2;M_5(x))$. Then, since $\al^5\eq 1$, their
respective parity-check matrices are
\begin{eqnarray*}
\cH^{(2)}(3,5,1,2)&=&\left(
\begin{array}{ccccc|ccccc|ccccc}
1&1&1&1&1&0&0&0&0&0&0&0&0&0&0\\
0&0&0&0&0&1&1&1&1&1&0&0&0&0&0\\
0&0&0&0&0&0&0&0&0&0&1&1&1&1&1\\
\hline
1&\al&\al^2&\al^3&\al^4&1&\al&\al^2&\al^3&\al^4&1&\al&\al^2&\al^3&\al^4\\
1&\al&\al^2&\al^3&\al^4&\al&\al^2&\al^3&\al^4&1&\al^2&\al^3&\al^4&1&\al\\
\end{array}
\right)\\
\end{eqnarray*}
\begin{eqnarray*}
\cH^{(2)}(5,3,1,2)&=&\left(
\begin{array}{ccc|ccc|ccc|ccc|ccc}
1&1&1&0&0&0&0&0&0&0&0&0&0&0&0\\
0&0&0&1&1&1&0&0&0&0&0&0&0&0&0\\
0&0&0&0&0&0&1&1&1&0&0&0&0&0&0\\
0&0&0&0&0&0&0&0&0&1&1&1&0&0&0\\
0&0&0&0&0&0&0&0&0&0&0&0&1&1&1\\
\hline
1&\al&\al^2&1&\al&\al^2&1&\al&\al^2&1&\al&\al^2&1&\al&\al^2\\
1&\al&\al^2&\al&\al^2&\al^3&\al^2&\al^3&\al^4&\al^3&\al^4&1&\al^4&1&\al\\
\end{array}
\right)\\
\end{eqnarray*}
}
\end{ex}

The following lemma is immediate:

\begin{lemma}
\label{lsimple}
{\em
The code $\C^{(2)}(m,n,1,1;f(x))$ given by
Construction~\ref{c3} is PMDS.
}
\end{lemma}

Comparing lemmas~\ref{th00} and~\ref{lsimple}, both
$\C(m,n,1,1;f(x))$ and $\C^{(2)}(m,n,1,1;g(x))$ are PMDS, where
$f(x)$ and $g(x)$ are either irreducible or have the form $M_p(x)$
for some prime number $p$. However,
the conditions on $\C^{(2)}(m,n,1,1;g(x))$ are less stringent. For
instance, if we consider the codes of Example~\ref{exs} for $M_p(x)$,
we can see that we need to consider at least $p\eq 17$ for  
$\C(3,5,1,1;M_p(x))$ and $\C(5,3,1,1;M_p(x))$, while we may take
$p\eq 5$ for $\C^{(2)}(3,5,1,1;M_p(x))$ and
$\C^{(2)}(5,3,1,1;M_p(x))$. Thus, although we are using a smaller
field or ring, 
the PMDS property is not lost. This is not the case for codes
$\C^{(2)}(m,n,1,2;f(x))$: we immediately see that the codes are not
(1;2)-erasure correcting (and hence are not PMDS). However, they are
(1;1,1)-erasure correcting, as stated in the following lemma: 

\begin{lemma}
\label{lsimple2}
{\em
The code $\C^{(2)}(m,n,1,2;f(x))$ given by
Construction~\ref{c3} is (1;1,1)-erasure correcting.
}
\end{lemma}

\pf Assume that we have two erasures in locations $j_0$ and $j_1$ of
row $i_0$ and two erasures in locations $\ell_0$ and $\ell_1$ of
row $i_1$, where $0\leq i_0<i_1\leq m-1$, $0\leq j_0<j_1\leq n-1$ and
$0\leq \ell_0<\ell_1\leq n-1$. Using the parity-check matrix
$\cH^{(2)}(m,n,1,2)$ as given in Construction~\ref{c3}, these four
erasures can be recovered if and only if 
$$
\det\left(
\begin{array}{cccc}
1&1&0&0\\
0&0&1&1\\
\al^{j_0}&\al^{j_1}&\al^{\ell_0}&\al^{\ell_1}\\
\al^{i_0+j_0}&\al^{i_0+j_1}&\al^{i_1+\ell_0}&\al^{i_1+\ell_1}\\
\end{array}
\right)
$$
is invertible. By row operations, we find out that this determinant
is invertible if and only if $1\xor \al^{j_1-j_0}$, $1\xor
\al^{\ell_1-\ell_0}$ and $1\xor \al^{i_1-i_0}$ are invertible. But
this is certainly the case
if $f(x)$ is irreducible or $f(x)\eq M_p(x)$, since $j_1-j_0$,
$\ell_1-\ell_0$ and $i_1-i_0$ are smaller than the exponent of $f(x)$.
\qed

Lemma~\ref{lsimple2} is important in applications. Let us compare it
with $\C(m,n,1,2;f(x))$ codes that are PMDS as given in
Subsection~\ref{cmn12}. For the sake of discussion, let us assume
that $\max\{m,n\}\leq 15$, a situation that covers some practical
applications. In the case of $\C(m,n,1,2;f(x))$, using
Table~\ref{t1}, if $n\eq 15$, we would need to operate on the field
$GF(2^{16})$. If we use a code $\C^{(2)}(m,n,1,2;f(x))$, we can take
the finite field $GF(2^4)$ as given by a primitive polynomial, which
has exponent 15. If we use $GF(2^5)$ with a primitive polynomial, we
can increase $m$ to 16, a value convenient in applications. If we use
rings generated by $M_p(x)$ and we want $m\leq 17$, $n\leq 17$, by
Table~\ref{t2}, we may use $p\eq 257$ for a PMDS code
$\C(m,n,1,2;M_{257}(x))$. If we just implement a (1;1,1)-erasure
correcting code for the same values of $m$ and $n$, we can do it with
a code $\C^{(2)}(m,n,1,2;M_{17}(x))$. 

Let us point out that Construction~\ref{c3} is closely related to Generalized
Concatenated (GC) codes~\cite{bz}\cite{z}. For descriptions of GC
codes, see also~\cite{b}\cite{du}\cite{wc} and the references
therein. Implementations of GC codes are given by two-level ECC
schemes~\cite{ah}\cite{c}\cite{pa}, later improved in the
two-level~\cite{hl}\cite{hapkt} and the multilevel~\cite{tk} Integrated
Interleaving schemes.

Using ideas similar to the 
ones of GC codes we can
extend Construction~\ref{c3} to codes that are 
$(1;\overbrace{1,1,\ldots,1}^s)$-erasure correcting (that 
we denote $\C^{(2)}(m,n,1,s;f(x))$)
as well as other combinations by using horizontal
and vertical codes, but for reasons of space we omit them here.
Moreover, as we will see in the next section, there is not much gain
for codes $\C^{(2)}(m,n,1,s;f(x))$ and $s\geq 3$ with respect to
codes $\C^{(2)}(m,n,1,2;f(x))$ in a mixed environment of catastrophic
failures and hard errors. 

\section{Probability of Data Loss After One Disk Failure}
\label{prob}
In this section, we assume that a catastrophic device failure has
occurred. We will make a number of assumptions and we will compute
the probability of data loss for the different schemes presented in
the paper as a function of the raw error probability $p$ (a parameter
that, as we have discussed in the Introduction, degrades with time
and with the number of writes for SSDs). Specifically, we will
compare $(1;2)$ PMDS codes and %$(\overbrace{2,2,\ldots,2}^s)$
(1;1,1)-erasure correcting codes, since both have the same redundancy
(but the former is implemented over a larger field). We assume that the
information in each SSD is stored in pages, where each
page has size 4K and there are eight 512B sectors per page. Further, 
we assume that each
sector is protected by a $t$-bit error-correcting code, 
like a BCH code (for instance, $t\eq 15$).
Each SSD device has $M$ pages. For example, if a device has size 32
G, it has 8 million pages. We assume that stripes are rows of pages
in an $m\times n$ block.

Since we assume that one of the $n$ devices has failed, if exactly
one hard error has occurred in at least three different stripes of an
$m$-stripe block, we will have data loss. If a $(1;2)$ PMDS code is
used, three hard errors in the same stripe will also cause data loss, while
if a  $(1;1,1)$-erasure correcting code is used,
two hard errors in the same stripe are enough to cause data loss.

As stated above, each codeword is in a BCH code with 512B information bytes (4096 bits). 
The BCH code can correct up to $t$ bit errors, so the redundancy is $13t$ bits,
giving 195 bits for $t\eq 15$. 

%There are eight codewords per page. 
We want to compute first the probability $P$ 
that a codeword cannot be decoded. This will occur each time $t+1$ or more
errors occur, and this event we assume is always detected either by the 
BCH code itself or by the CRC.
Therefore, we have:

\begin{eqnarray*}
P & = & \sum_{i=t+1}^{4096+13t}\,{{4096+13t}\choose i}\,p^i\,(1-p)^{4096+13t-i}
\end{eqnarray*}

Thus, since there are 8 sectors per page, 
the probability that in a page at least a codeword is not corrected
(i.e., a hard error) is 

\begin{eqnarray*}
P_{\rm H} & = & \sum_{i=1}^8 {8\choose i}P^i\,(1-P)^{8-i}\;\eq\; 1-(1-P)^8.
\end{eqnarray*}

The probability of exactly one hard error in a stripe is
\begin{eqnarray*}
%P_{{\rm HR}\geq 1} & = & (n-1) P_{\rm H}(1-P_{\rm H})^{n-2}\\
P_{{\rm HR}=1} & = & (n-1) P_{\rm H}(1-P_{\rm H})^{n-2}\\
\end{eqnarray*}

The probability of more than $j$ hard errors in a stripe is
\begin{eqnarray*}
P_{{\rm HR}> j} & = & \sum_{i=j+1}^{n-1} {{n-1}\choose i}\left(P_{\rm H}\right)^i\,\,
(1-P_{\rm H})^{n-i-1}\\
 \end{eqnarray*}

%The probability of three or more hard errors in a stripe is
%\begin{eqnarray*}
%P_{{\rm HR}> 2} & = & \sum_{i=3}^{n-1} {{n-1}\choose i}\left(P_{\rm H}\right)^i\,\,
%(1-P_{\rm H})^{n-i-1}\\
% \end{eqnarray*}

The probability of exactly one hard error in 
at least three of the $m$ stripes in a block is then

\begin{eqnarray*}
P_{{\rm S}\,m,1,3} & = &\sum_{i=3}^{m} {{m}\choose i}\left(P_{{\rm HR}=1}\right)^i\,\,
(1-P_H)^{(n-1)(m-i)}
\end{eqnarray*}

The probability of at least $j+1$ hard errors 
in any of the $m$ stripes of the  block is
\begin{eqnarray*}
P_{{\rm S}\,m,j+1,1} & = & mP_{{\rm HR}>j}
%\sum_{i=1}^{m} {m\choose i}\left(P_{{\rm
%HR}>1}\right)^i\,\,(1-P_{{\rm HR}>1})^{m-i}\\ 
\end{eqnarray*}

The probability of data loss in an $m$-stripe block of a 
(1;1,1)-erasure correcting code is then given by
\begin{eqnarray*}
P_{\rm S\,(1;1,1)EC} & = & P_{{\rm S}\,m,1,3} +P_{{\rm S}\,m,2,1},
\end{eqnarray*}
while the probability of data loss in an $m$-stripe block of a (1,2)
PMDS code is given by
\begin{eqnarray*}
P_{\rm S\,(1;2)PMDS} & = & P_{{\rm S}\,m,1,3} +P_{{\rm S}\,m,3,1}.
\end{eqnarray*}

We can now compute the probability of data loss for both a
(1;1,1)-erasure correcting code and a (1;2) PMDS code. That will occur each
time at least one $m$-stripe block has experienced data loss. Thus,
since we had assumed that there are $M$ pages per device and that
each block has $m$ stripes, there are $M/m$ blocks (for 32G SSDs,
and $m\eq 16$, $M/m\eq 500,000$), we obtain for a (1;1,1)-erasure correcting code, 

\begin{eqnarray*}
P_{\rm DL\,(1;1,1)EC} & = & \sum_{i=1}^{M/m} {M/m\choose i}\left(P_{\rm
S\,(1;1,1)EC}\right)^i\,\,(1-P_{\rm S\,(1;1,1)EC})^{(M/m)-i}\\ 
\end{eqnarray*}
and for a (1;2) PMDS code, 
\begin{eqnarray*}
P_{\rm DL\,(1;2)PMDS} & = & \sum_{i=1}^{M/m} {M/m\choose i}\left(P_{\rm
S\,(1;2)PMDS}\right)^i\,\,(1-P_{\rm S\,(1;2)PMDS})^{(M/m)-i}\\ 
\end{eqnarray*}

Looking at the probabilities of data loss in an $m$-stripe block for
$m\eq 16$ and 32G devices in Table~\ref{t3}, we 
can see that in general $P_{\rm S\,(1;1,1)EC}$ is dominated by
$P_{{\rm S}\,m,2,1}$ (i.e., $P_{\rm S\,(1;1,1)EC}\approx P_{{\rm
S}\,m,2,1}$), while $P_{\rm S\,(1;2)PMDS}$ is dominated by $P_{{\rm
S}\,m,1,3}$ (i.e., $P_{\rm S\,(1;2)PMDS}\approx P_{{\rm S}\,m,1,3}$).
For that reason, by increasing $s$, the probability of data loss of a
$(1;\overbrace{1,1,\ldots,1}^s)$-erasure correcting code is basically
the same for any $s\geq 2$ when a whole device has failed. In
particular, for $s\eq m$, we have RAID 6. Of course RAID 6 can
tolerate a second device failure, but any hard error in the case of
two device failures will cause data loss.

Another conclusion from Table~\ref{t3} is the advantage of using a
(1;2) PMDS code over a (1;1,1)-erasure correcting code, both codes
having the same number of parity entries. As stated in the
Introduction, as the system ages, the bit error probability $p$
degrades. So, a natural question is, if we are monitoring $p$, which
value allows us a reasonable expectation of not experiencing data
loss? For instance, when we reach $p\eq .0007$, according to
Table~\ref{t3}, the probability of miscorrection in case a device
fails is 7.8E-5. This may be viewed as, less than one in ten thousand
systems will have data loss provided a device has failed, which may be
acceptable (depending on the application). However, if we used a
(1;2) PMDS code, when $p\eq .0008$, 
the probability of data loss is 6.3E-6, more than an order of
magnitude better than the (1;1,1)-erasure correcting code. So, the
system is more reliable and allows further degradation of the
parameter $p$, increasing its lifetime.

\begin{table}
\begin{center}
\begin{tabular}{|c|c|c|c|c|c|c|c|c|c|c|}
\hline
$p$&.0001&.0002&.0003&.0004&.0005&.0006&.0007&.0008&.0009&.001\\
\hline\hline
{\scriptsize $P$}&{\scriptsize 4.1E-20}&{\scriptsize
1.8E-15}&{\scriptsize 7.9E-13}&{\scriptsize 5.3E-11}&{\scriptsize
1.3E-9}&{\scriptsize 1.6E-8}&{\scriptsize 1.2E-7}&{\scriptsize
7.0E-7}&{\scriptsize 3.1E-6}&{\scriptsize 1.1E-5}\\   
\hline
{\scriptsize $P_{\rm H}$}&{\scriptsize 3.3E-19}&{\scriptsize
1.4E-14}&{\scriptsize 6.3E-12}&{\scriptsize 7.9E-13}&{\scriptsize
1.0E-8}&{\scriptsize 1.3E-7}&{\scriptsize 9.9E-7}&{\scriptsize
5.6E-6}&{\scriptsize 2.5E-5}&{\scriptsize 9.0E-5}\\   
\hline\hline
{\scriptsize $P_{{\rm S}\,16,1,3}$}&{\scriptsize 2.5E-51}&{\scriptsize
2.1E-37}&{\scriptsize 1.8E-29}&{\scriptsize 5.3E-24}&{\scriptsize
7.2E-20}&{\scriptsize 1.4E-16}&{\scriptsize 6.8E-14}&{\scriptsize
1.3E-11}&{\scriptsize 1.1E-9}&{\scriptsize 5.2E-8}\\   
\hline
{\scriptsize $P_{{\rm S}\,16,2,1}$}&{\scriptsize 5.3E-35}&{\scriptsize
3.3E-26}&{\scriptsize 6.4E-21}&{\scriptsize 2.9E-17}&{\scriptsize
1.6E-14}&{\scriptsize 2.5E-12}&{\scriptsize 1.6E-10}&{\scriptsize
5.1E-9}&{\scriptsize 1.0E-9}&{\scriptsize 1.3E-6}\\   
\hline
{\scriptsize $P_{{\rm S}\,16,3,1}$}&{\scriptsize 5.7E-54}&{\scriptsize
4.8E-40}&{\scriptsize 4.1E-32}&{\scriptsize 1.2E-26}&{\scriptsize
3.1E-19}&{\scriptsize 3.1E-19}&{\scriptsize 1.6E-16}&{\scriptsize
2.9E-14}&{\scriptsize 2.5E-12}&{\scriptsize 1.2E-10}\\   
\hline\hline
{\scriptsize $P_{\rm S\,(1;1,1)EC}$}&{\scriptsize 1.7E-35}&{\scriptsize
3.3E-26}&{\scriptsize 6.4E-21}&{\scriptsize 2.9E-17}&{\scriptsize
1.6E-14}&{\scriptsize 2.5E-12}&{\scriptsize 1.6E-10}&{\scriptsize
5.1E-9}&{\scriptsize 1.0E-7}&{\scriptsize 1.4E-6}\\   
\hline
{\scriptsize $P_{\rm S\,(1;2)PMDS}$}&{\scriptsize
2.5E-51}&{\scriptsize 2.1E-37}&{\scriptsize 1.8E-29}&{\scriptsize
5.3E-24}&{\scriptsize 7.2E-20}&{\scriptsize 1.4E-16}&{\scriptsize
6.8E-14}&{\scriptsize 1.3E-11}&{\scriptsize 1.1E-9}&{\scriptsize
5.2E-8}\\    
\hline\hline
{\scriptsize $P_{\rm DL\,(1;1,1)EC} $}&{\scriptsize
8.6E-30}&{\scriptsize 1.7E-20}&{\scriptsize 3.2E-15}&{\scriptsize
1.4E-11}&{\scriptsize 8.2E-9}&{\scriptsize 1.3E-6}&{\scriptsize
7.8E-5}&{\scriptsize 2.5E-3}&{\scriptsize .05}&{\scriptsize .5}\\  
\hline
{\scriptsize $P_{\rm DL\,(1;2)PMDS}$}&{\scriptsize
1.2E-45}&{\scriptsize 1.1E-31}&{\scriptsize 8.9E-24}&{\scriptsize
2.7E-18}&{\scriptsize 3.6E-14}&{\scriptsize 6.9E-11}&{\scriptsize
3.4E-8}&{\scriptsize 6.3E-6}&{\scriptsize 5.4E-4}&{\scriptsize .026}\\   
\hline
\end{tabular}
\caption{Probabilities of data loss for (1;1,1)-erasure correcting
codes and (1;2) PMDS codes for different values of bit error
probability $p$ in the presence of a catastrophic device failure
}
\label{t3}
\end{center}
\end{table}

\section{Conclusions}
We have presented two constructions of codes that are suitable for a
flash array type of architecture, in which hard errors co-exist with
catastrophic device failures. We have presented specific codes that
are useful in applications. Necessary and sufficient conditions for
codes satisfying an optimality criterion were given. 

\appendix
\section{Appendix}

\begin{lemma}
{\em
Let $\ga_0,\ga_1,\ldots ,\ga_{s-1}$ be distinct elements in a field
or ring of characteristic 2.
Consider the $s\times s$ matrix 
\begin{eqnarray*}
\Gamma&=&
\left(\begin{array}{cccc}
\ga_0&\ga_1&\ldots &\ga_{s-1}\\
\ga_0^2&\ga_1^2&\ldots &\ga_{s-1}^2\\
\ga_0^4&\ga_1^4&\ldots &\ga_{s-1}^4\\
\vdots &\vdots &\ddots &\vdots \\
\ga_0^{2^{s-1}}&\ga_1^{2^{s-1}}&\ldots &\ga_{s-1}^{2^{s-1}}\\
\end{array}\right)
\end{eqnarray*}
Then, 
\begin{eqnarray*}
\det\Gamma&=&\prod_{S\subseteq
\{\ga_0,\ga_1\ldots,\ga_{s-1}\}}\,\bigoplus_{i\in S}\,\ga_i
\end{eqnarray*}
}
\end{lemma}

\pf We will do induction on $s$. The result is certainly true for
$s\eq 1$. Consider the determinant of the matrix obtained by
replacing $\ga_0$ in the first column of $\Gamma$ by $x$, i.e., 
\begin{eqnarray*}
h(x)&=&
\det\left(\begin{array}{cccc}
x&\ga_1&\ldots &\ga_{s-1}\\
x^2&\ga_1^2&\ldots &\ga_{s-1}^2\\
x^4&\ga_1^4&\ldots &\ga_{s-1}^4\\
\vdots &\vdots &\ddots &\vdots \\
x^{2^{s-1}}&\ga_1^{2^{s-1}}&\ldots &\ga_{s-1}^{2^{s-1}}\\
\end{array}\right)
\end{eqnarray*}
Since $h(x)$ has degree $2^{s-1}$, it has at most $2^{s-1}$ zeros.
Notice that if $S$ is one of the $2^{s-1}$ subsets of
$\{\ga_1,\ga_2\ldots,\ga_{s-1}\}$ (including the empty subset), then 
$\bigoplus_{i\in S}\,\ga_i$ is a zero of $h(x)$ (the element 0
corresponding to the empty set), due to the linearity of the square
operation in a field of characteristic 2. Therefore, we can write, 
\begin{eqnarray*}
h(x)&=&C\prod_{S\subseteq
\{\ga_1,\ga_2\ldots,\ga_{s-1}\}}\,\left(x+ \bigoplus_{i\in
S}\,\ga_i\right), 
\end{eqnarray*}
where 
\begin{eqnarray*}
C&=&\det
\left(\begin{array}{cccc}
\ga_1&\ga_2&\ldots &\ga_{s-1}\\
\ga_1^2&\ga_2^2&\ldots &\ga_{s-1}^2\\
\ga_1^4&\ga_2^4&\ldots &\ga_{s-1}^4\\
\vdots &\vdots &\ddots &\vdots \\
\ga_1^{2^{s-2}}&\ga_2^{2^{s-2}}&\ldots &\ga_{s-2}^{2^{s-2}}\\
\end{array}\right)
\end{eqnarray*}
The result follows from the fact that $h(\ga_0)\eq \det(\Gamma)$ and
by induction on the expression of $C$ above.\qed

\end{document}